\documentclass{ieeeaccess}

\usepackage{cite}
\usepackage{amsmath,amssymb,amsfonts}
\usepackage{algorithmic}
\usepackage{graphicx}
\usepackage{textcomp}
\usepackage{xcolor}
\usepackage{listings}

\usepackage{bm}
\makeatletter
\AtBeginDocument{\DeclareMathVersion{bold}
\SetSymbolFont{operators}{bold}{T1}{times}{b}{n}
\SetSymbolFont{NewLetters}{bold}{T1}{times}{b}{it}
\SetMathAlphabet{\mathrm}{bold}{T1}{times}{b}{n}
\SetMathAlphabet{\mathit}{bold}{T1}{times}{b}{it}
\SetMathAlphabet{\mathbf}{bold}{T1}{times}{b}{n}
\SetMathAlphabet{\mathtt}{bold}{OT1}{pcr}{b}{n}
\SetSymbolFont{symbols}{bold}{OMS}{cmsy}{b}{n}
\renewcommand\boldmath{\@nomath\boldmath\mathversion{bold}}}
\makeatother

\def\BibTeX{{\rm B\kern-.05em{\sc i\kern-.025em b}\kern-.08em
    T\kern-.1667em\lower.7ex\hbox{E}\kern-.125emX}}

\lstdefinestyle{yamlfragment}{
  basicstyle=\ttfamily\scriptsize,
  breaklines=true,
  columns=fullflexible,
  keepspaces=true,
  frame=single
}

\usepackage{xurl}
\usepackage[hidelinks]{hyperref}

\begin{document}
\bstctlcite{IEEEexample:BSTcontrol}

\history{Received 14 June 2026, accepted 27 July 2026, date of publication 4 August 2026, date of current version 13 August 2026.}
\doi{10.1109/ACCESS.2026.3720448}
\vol{14}
\year{2026}

\title{Agent-Native Applications: A New Application Paradigm}

\author{\uppercase{David Luo}\authorrefmark{1}, AND
\uppercase{Alberto Leon-Garcia}\authorrefmark{1}, \IEEEmembership{Life Fellow, IEEE}}

\address[1]{Department of Electrical and Computer Engineering, University of Toronto, Toronto, ON, Canada}

\tfootnote{This is the authors' accepted manuscript of an article
published in \textit{IEEE Access}, vol.~14,
pp.~121508--121528, 2026.
The version of record is available at
\url{https://doi.org/10.1109/ACCESS.2026.3720448}.}

\markboth
{D. Luo, A. Leon-Garcia: Agent-Native Applications}
{D. Luo, A. Leon-Garcia: Agent-Native Applications}

\corresp{Corresponding author: David Luo (e-mail: david.luo@mail.utoronto.ca).}

\begin{abstract}
Traditional application architectures assume behavioral logic authored in advance, leaving reachable behavior largely bounded by explicit code and workflow rules. Large language model (LLM)-based agents challenge this assumption by enabling runtime reasoning and autonomous action to become part of application behavior, allowing applications to address situations not enumerated at design time. Extending this pattern, this paper identifies a new application paradigm in which LLM-based agents serve as central reasoning and action components responsible for the application's core logic and, where permitted, for adapting the application graph itself at runtime. We call these \emph{agent-native} applications. While such applications significantly expand their possible behavioral space beyond explicit code and workflow rules, they also face a major control problem in which useful agentic reasoning should be preserved while application behavior should remain within a permissible space. We therefore propose an architectural model that represents the application as a portable graph of agents, tools, data sources, and human-in-the-loop (HITL) checkpoints, and encodes the application's permitted behavior as a behavioral envelope within a declarative application specification. At runtime, an application orchestrator serves as the control plane that coordinates tasks and governs how the graph and its permissions evolve, while an agent mesh serves as the data plane that mediates policy-relevant interactions and produces audit events. We then discuss the trust layer that makes agent-native applications governable and the supporting foundations required for practical operation. Two use cases illustrate the architecture, while Agent-Native Runtime (ANR) demonstrates selected core mechanisms in an executable prototype.
\end{abstract}

\begin{keywords}
Agent-native applications, application paradigm, LLM agents, multi-agent systems, software architecture, agent orchestration, declarative specification, policy enforcement, human-in-the-loop governance.
\end{keywords}

\titlepgskip=-21pt

\maketitle

\section{Introduction}
\label{sec:introduction}

\PARstart{A}{pplication} architecture is shifting from systems with pre-authored behavior toward systems whose behavior is shaped at runtime by large language model (LLM) agents. Historically, successive architectural paradigms have changed how applications are decomposed and deployed, yet the logic that determines how an application responds to each input has remained largely authored in advance through deterministic code and workflow rules. Such a design works well when desired behavior can be specified ahead of time, but it is significantly limited when tasks require contextual judgment, flexible action, and adaptation to situations not enumerated at design time. LLM-based agents provide a promising mechanism for overcoming this limitation by bringing general-purpose runtime reasoning into application behavior. They therefore make possible a new application paradigm in which runtime agentic reasoning becomes a core operating mechanism of the application itself.

This shift is becoming plausible because recent LLM agents increasingly exhibit capabilities needed to perform core application-level roles, including persistence over long-horizon multi-step objectives and coordination across specialized agents. Long-running coding and scientific-computing examples show agents pursuing task objectives through iterative loops of planning, tool use, observation, validation, and repair over extended execution, including a 25-hour Codex run~\cite{openai2026codex_long_horizon} and multi-day Claude Code workflows guided by persistent memory and test oracles~\cite{anthropic2026long_running_claude_scientific}. Multi-agent experiments likewise show that agent teams can coordinate on substantial software tasks through parallel workers in specialized roles, organized through task assignment and judge or synchronization mechanisms, including compiler-building~\cite{anthropic2026parallel_claudes_c_compiler} and browser-scale coding experiments~\cite{cursor2026scaling_agents}. Although agents remain stochastic and imperfect, these developments suggest that agent capabilities are approaching the threshold at which they can shoulder architectural roles previously held by deterministic application logic or workflow engines.

The increasing capability of LLM agents shifts the architectural question from how to add agents to existing applications toward how to design applications whose core behavior is organized around agents. Such applications are distinct from chatbot, assistant, or artificial intelligence (AI) feature patterns appended to traditional software. Instead, they are applications in which agents are central constituents of the architecture, responsible for the application's core logic and coordination. Even the application graph can be reshaped dynamically through agentic reasoning, allowing the application's composition and structure to adapt at runtime. This is the sense in which the paper treats agent-native applications as a new \emph{general class} of applications.

The same generality, stochasticity, and dynamic runtime reasoning that make agents architecturally significant also create the central control problem of this paradigm. While deterministic programs can also exhibit erratic behavior through failure, bugs, or compromise, their normal reachable behavior is largely bounded by explicit control flow, state transitions, and interfaces. LLM-based agents instead propose actions from a probabilistic, context-dependent space whose long tail can include unsafe tool use, unintended data disclosure, and other behavior that cannot be easily anticipated. These risks become central when agents take on a core role in applications and their outputs can trigger application actions with meaningful consequences. Applications that place agents at the center therefore need mechanisms that bind realized behavior to a permissible space without collapsing agentic flexibility toward fixed workflow execution.

We call this emerging application class the \emph{agent-native} application paradigm. An application is agent-native when LLM-based agents are central reasoning and action components, responsible for the application's core logic and coordination and, where permitted, for adapting the application graph itself at runtime. We use \emph{agentic} more broadly for applications in which LLM-based agents take any functional role, including auxiliary or narrowly scoped assistant roles. \emph{Agent-native} is reserved for applications whose core behavior is organized around runtime agentic reasoning.\footnote{The term \emph{agent-native} has been used previously by others with varied meanings. We fix it here to the definition in the main text. Throughout this paper, \emph{agent} refers to an LLM-based component capable of autonomous reasoning and tool calls. \emph{Software} refers to traditional deterministic programs, compiled or interpreted code with explicitly pre-programmed logic. \emph{Application} refers to the larger system that may contain agents, traditional software, data sources, tools, and human-in-the-loop checkpoints.} Agent-native computing does not replace deterministic software. Workloads requiring exact deterministic behavior, or whose per-decision value does not justify LLM inference costs, will likely continue to rely on traditional software as tools or services within the application graph.

In agent-native applications, the central challenge is to constrain what the application may realize while retaining the agents' ability to complete useful tasks autonomously. We therefore propose an architectural model for specifying, bounding, and enforcing agent-native application behavior at the application level. The model centers on a \emph{behavioral envelope}, which defines the permissible behavior space of the application and is encoded in a declarative application specification. At runtime, an application orchestrator coordinates tasks and governs how the graph evolves within this envelope, while an agent mesh mediates policy-relevant interactions on the interaction path and produces oversight and audit events. Together, these mechanisms preserve the flexibility that makes agentic reasoning useful while constraining realized behavior independently of any one agent framework.

The paper's main contribution is to identify agent-native applications as an emerging application paradigm and to propose how their behavior can be specified, bounded, and enforced at the application-architecture level. Specifically, we:
\begin{itemize}
\item Define the agent-native paradigm as encompassing a general class of applications structured around runtime agentic reasoning as core application logic.
\item Develop a portable graph-based application model whose behavioral envelope, encoded in a declarative application specification, is enforced at runtime by a control-plane application orchestrator and a data-plane agent mesh.
\item Specify the trust layer and supporting foundations for agent-native applications, and illustrate the paradigm through two cross-domain use cases.
\item Implement the Agent-Native Runtime (ANR), a companion prototype demonstrating declarative specification, behavioral envelope enforcement, human oversight, and bounded graph evolution in an executable system.
\end{itemize}

\begin{figure*}[t]
\centering
\includegraphics[width=\textwidth,height=0.4\textheight,keepaspectratio]{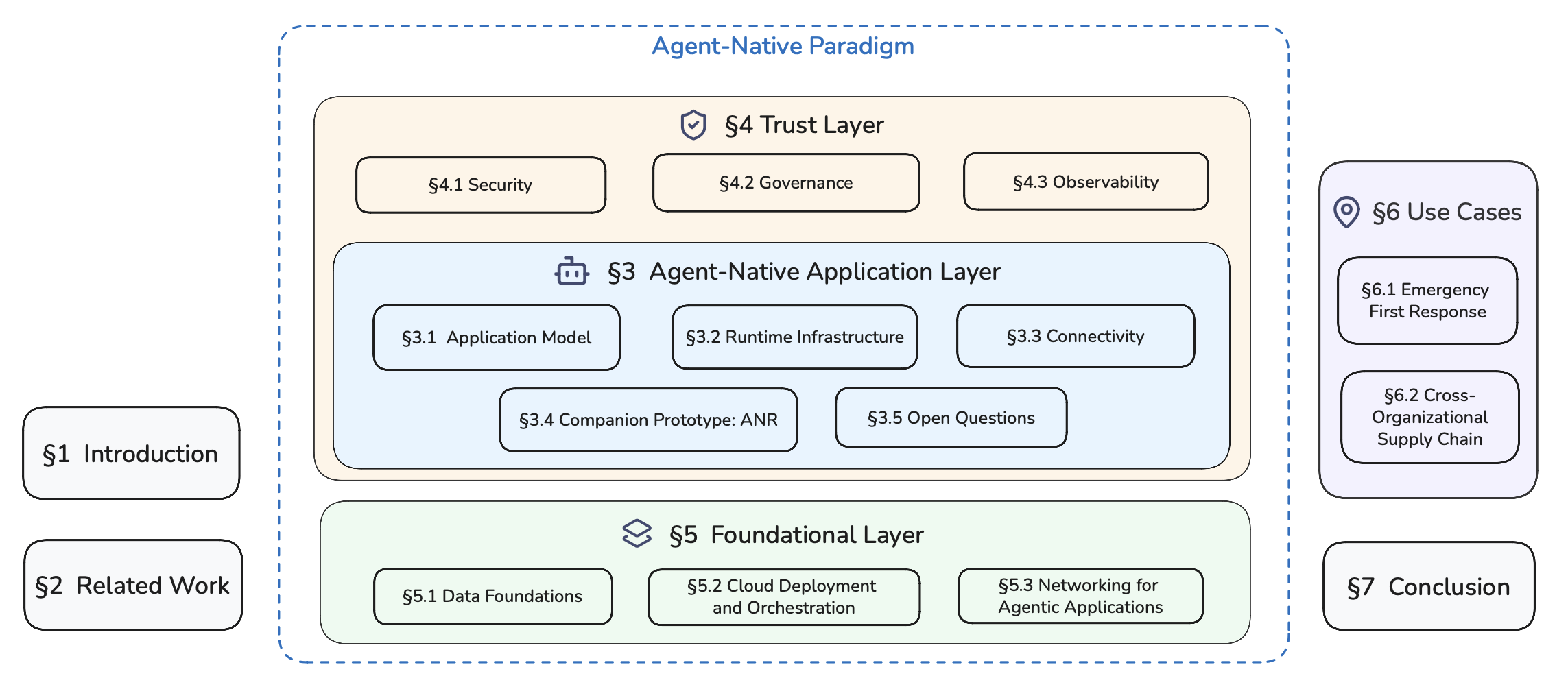}
\caption{Roadmap of the paper and the main architectural layers of the agent-native paradigm.}
\label{fig:sections-roadmap}
\end{figure*}

Fig.~\ref{fig:sections-roadmap} presents the paper's organization. Section~\ref{sec:related-work} reviews prior work that informs the agent-native paradigm. Section~\ref{sec:agent-native-layer} develops the agent-native application layer. It defines the graph-based application model and behavioral envelope, explains the declarative application specification that encodes the envelope, and describes runtime enforcement through the application orchestrator and agent mesh. It also addresses the connectivity mechanisms through which application components interact. Section~\ref{sec:trust-layer} develops the trust layer that makes runtime agentic reasoning governable across security, governance, and observability. Section~\ref{sec:foundational-layer} turns to the foundations on which the paradigm runs, covering data organization, deployment across hybrid and edge cloud environments, and network support for agentic communication patterns. Section~\ref{sec:use-cases} grounds the agent-native paradigm in specific application scenarios.

\section{Related Work}
\label{sec:related-work}

Several bodies of prior work help build toward the agent-native application paradigm. Pre-LLM agent systems developed conceptual foundations for autonomous coordination, while post-LLM orchestration frameworks have demonstrated agentic and multi-agent collaboration. Other works introduce standardized syntax and declarative means to specify agentic systems and the constraints that govern them. Meanwhile, emerging communication protocols and Internet of Agents proposals address interoperability across agentic entities. Finally, a further line of work moves agent safety and governance from framework-local guidance toward external runtime enforcement around agent execution.

The goal of composing autonomous components into collaborative applications predates LLM agents. Pre-LLM multi-agent systems were studied as an architectural response to problems where specialization, distribution, or negotiation between autonomous actors improved on a centralized design. Belief-Desire-Intention (BDI) architectures formalized agent reasoning in terms of beliefs, desires, and intentions, providing structured models of deliberation and action selection~\cite{rao1995bdi}. Foundation for Intelligent Physical Agents (FIPA) agent communication standards, including the FIPA Agent Communication Language (FIPA-ACL), established structured message exchange for interoperability across agents~\cite{du2025agent_communication}. Autonomic computing established a monitor-analyze-plan-execute control-loop pattern over a shared knowledge model (MAPE-K) for self-adaptive systems~\cite{kephart2003autonomic}. These early lines of work provide conceptual precedents for agent reasoning, inter-agent communication, and runtime adaptation. Pre-LLM agents, however, were typically bound to symbolic plans, domain-specific rules, or narrow learned models, limiting adaptation to situations represented within the system's design assumptions.

LLM-based agents fundamentally change the architectural significance of agent composition, as they enable general open-ended language-based reasoning about novel situations. Recent LLM-based orchestration frameworks demonstrate such capability at the programming level. Frameworks such as AutoGen~\cite{wu2023autogen}, LangGraph~\cite{langgraph_framework}, CrewAI~\cite{crewai_framework}, and the OpenAI Agents SDK~\cite{openai_agents_sdk} enable developers to build applications from LLM agents, tools, memory, and workflows, including multi-agent systems. Their unit of abstraction, however, remains a framework-specific program where the agentic system is expressed and executed as SDK code or framework-specific configuration. An emergent direction is to instead treat the agentic application specification as a portable architectural artifact specified independently of any one programming framework.

Transitioning from framework-specific construction to interoperable application artifacts requires a standardized way to declare an application before translating that declaration into a running system. A number of recent proposals address how declarative specification can be applied at different scopes within agentic systems. Cao et al.~\cite{cao2026auton} separate a declarative, language-agnostic specification for an agent from its runtime substrate at the individual-agent level. Daunis~\cite{daunis2025declarative_agent_workflows} introduces a domain-specific language for agent workflow orchestration that compiles to a backend-agnostic representation, allowing the same workflow specification to execute across multiple host languages. At the knowledge-context layer, Mouzouni~\cite{mouzouni2026context_kubernetes} develops an explicit Kubernetes analogy with YAML-based manifests, a reconciliation loop, and a permission model for delivering enterprise knowledge to agents under governance constraints. Safety and governance domain-specific languages (DSLs) such as AgentSpec and behavioral contracts similarly express constraints over agent behavior and runtime enforcement~\cite{wang2025agentspec, bhardwaj2026agent_behavioral_contracts}.

Another strand of work approaches agentic computing from the perspective of protocols and ecosystems. With many new agentic protocols such as the Model Context Protocol (MCP) and Agent2Agent (A2A) proposed in recent years, agent protocol surveys~\cite{yang2025survey_agent_protocols, kong2025agent_communication} classify emerging mechanisms for agent-to-tool and agent-to-agent interoperability, emphasizing that standardized protocols are needed to overcome fragmentation among agent systems. Moreover, Internet of Agents studies frame large-scale agent collaboration as an ecosystem and infrastructure problem, spanning dynamic discovery, communication, and collaborative orchestration across heterogeneous agents~\cite{cisco2025internet_agents, wang2026internet_of_agents}. Some argue that agent communication requires stack-level support for both structured exchange and semantic agreement, beyond conventional data-delivery protocols~\cite{fleming2025layered_protocol_iot_agents}.

Recent research and runtime systems also illustrate a move from framework-local guardrails toward external enforcement around agent execution. AgentBound introduces declarative access-control policies and a policy enforcement engine for securing MCP servers~\cite{buehler2026agentbound}, while Sandlock confines agent-executed code through unprivileged Linux primitives for filesystem, network, IPC, and syscall policies~\cite{wang2026sandlock}. NVIDIA OpenShell allows running autonomous agents in sandboxed environments with kernel-level isolation and declarative YAML policies that constrain access to local files, credentials, external networks, and inference routing~\cite{nvidia_openshell_overview}. These works move agent safety from prompt-level guidance toward independently enforceable runtime mediation. Their emphasis, however, is primarily implementation-level containment using particular runtimes to confine agent processes or tool servers and their access to host resources. This paper's agent-native model operates at a different level of abstraction. It does not prescribe specific sandboxing or isolation mechanisms. Instead, it focuses on the architectural object to be specified and governed, including the agent-native application graph and the behavioral envelope that constrains its actions and permitted evolution.

Taken together, these prior works establish many ingredients for agent-native applications, from autonomous coordination and agentic programming frameworks to declarative specification and runtime enforcement. A gap remains, however, in conceptualizing the agent-native application itself as the primary architectural object for a general class of applications. This paper addresses the gap by formulating agent-native applications as an emerging general class whose core behavior is organized around runtime agentic reasoning. Unlike framework-specific implementations or systems focused on individual-agent containment or policy, our proposed model addresses the application architecture as a whole, with a general model that supports multi-agent composition and application-level orchestration. Its target is a portable application-level specification for an agent-native application graph, together with the behavioral envelope that bounds how that graph behaves and evolves.

\section{Agent-Native Application Layer}
\label{sec:agent-native-layer}

This section develops the agent-native application as an architectural object. If agent-native applications are to be portable and governable, they must be described at the application layer where composition, runtime permissions, and permitted graph evolution can be specified. The standardization target is therefore the application graph and its cross-component permissions, human oversight requirements, and graph-evolution limits, which determine which runtime behaviors may be realized while remaining independent of any individual agent, tool protocol, or orchestration framework.

The core model links the application graph, declarative application specification, behavioral envelope, and runtime enforcement. The application graph represents the composition of agents, tools, data sources, and human-in-the-loop (HITL) checkpoints. A declarative application specification defines the graph's initial structure and encodes a behavioral envelope over what the application may realize. A conforming runtime instantiates or binds the running graph from that specification. Runtime enforcement is divided between an agent mesh that mediates policy-relevant interactions on the data path and an application orchestrator that governs graph-level coordination and adaptation. This architecture constrains the application's behavior without relying on any individual agent's self-restraint.

\begin{figure}[t]
\centering
\includegraphics[width=\columnwidth]{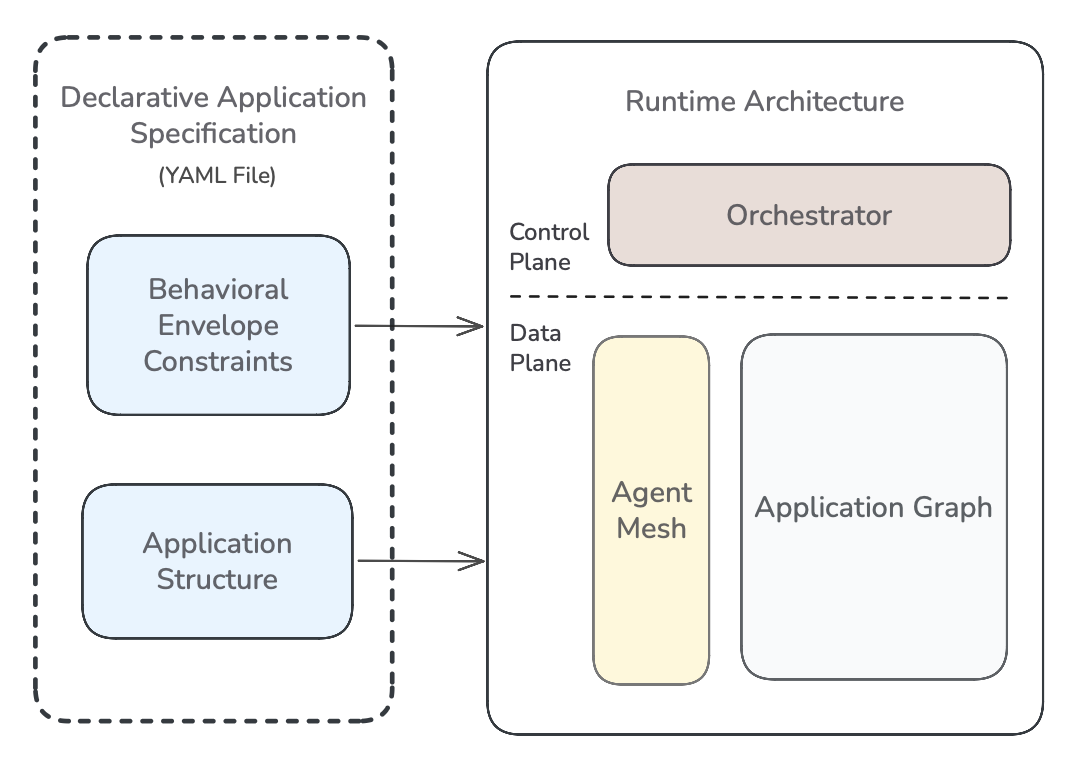}
\caption{Relationship among the declarative application specification, behavioral envelope, running application graph, agent mesh, and application orchestrator. The specification defines both the application structure and its permitted behavior. The runtime instantiates the application graph and enforces the encoded envelope through interaction-path mediation by the agent mesh and graph-level control by the application orchestrator.}
\label{fig:spec-envelope-runtime}
\end{figure}

Fig.~\ref{fig:spec-envelope-runtime} summarizes these relationships, which are developed in detail in the following sections. Section~\ref{sec:application-model} defines the graph-based application model, the declarative application specification, and the behavioral envelope encoded within it. Section~\ref{sec:runtime-infrastructure} introduces the application orchestrator and agent mesh as the control and data planes of runtime enforcement. Section~\ref{sec:connectivity} addresses the interaction contracts and protocols through which components communicate. Section~\ref{sec:anr-prototype} describes Agent-Native Runtime (ANR) as a focused companion prototype. Lastly, Section~\ref{sec:agent-native-open-questions} identifies open questions in realizing the agent-native application layer.

\subsection{Application Model}
\label{sec:application-model}

\begin{figure*}[t]
\centering
\includegraphics[width=\textwidth,height=0.4\textheight,keepaspectratio]{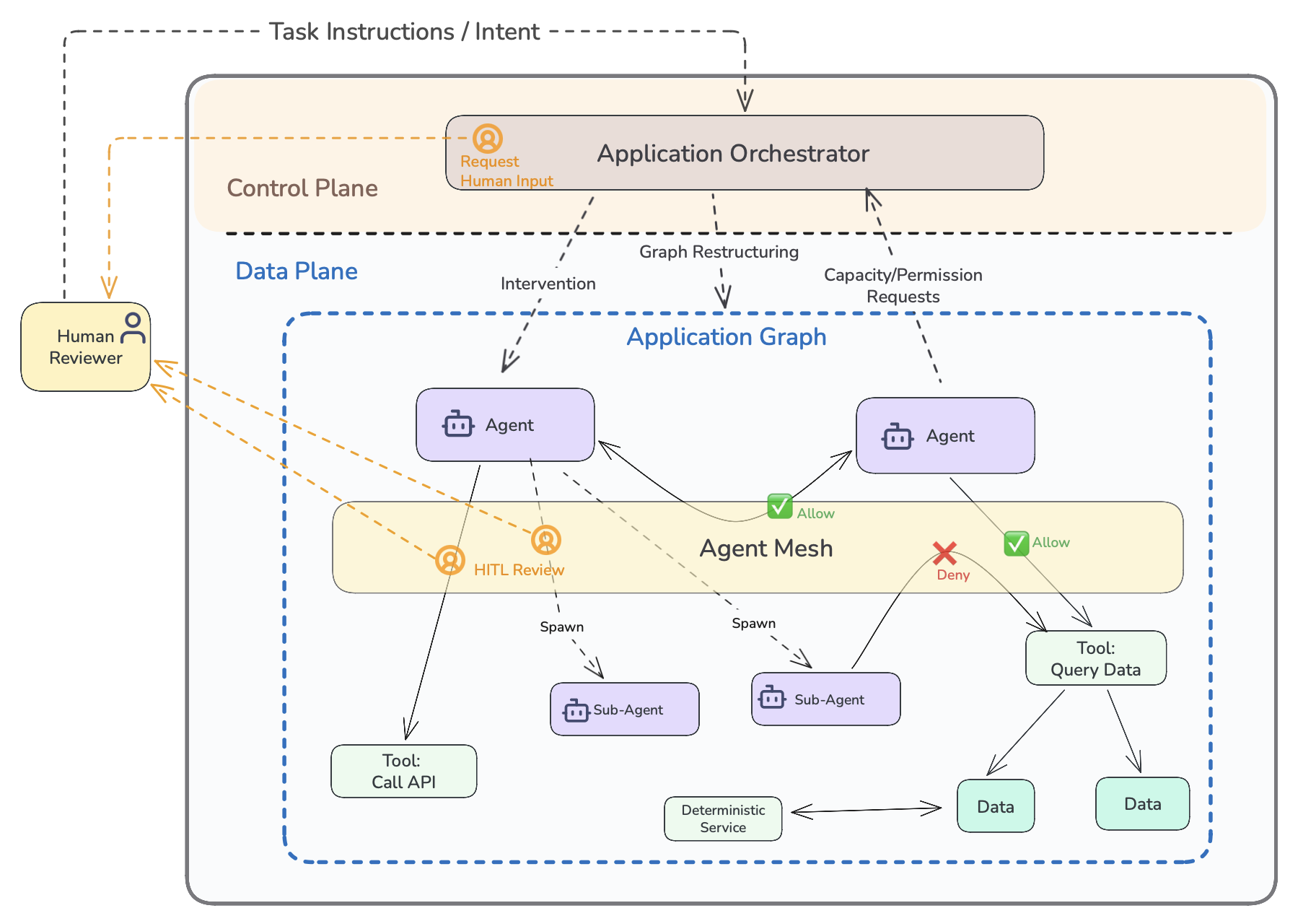}
\caption{Runtime anatomy of an agent-native application. The outer runtime boundary represents the behavioral envelope that bounds both the control-plane application orchestrator and the data-plane application graph. The agent mesh mediates governed interactions, including allow, deny, and HITL review outcomes, while the orchestrator handles graph restructuring, interventions, and permission requests.}
\label{fig:application-architecture}
\end{figure*}

\subsubsection{Agent-Native Application as Graph-Based Composition}

We treat the composition of an agent-native application as a graph, situated within the broader runtime anatomy illustrated in Fig.~\ref{fig:application-architecture}. The graph model serves two purposes. First, it represents the application's compositional structure as nodes interacting via edges. Second, it exposes the interaction points at which runtime agentic reasoning must be governed. Given that agents occupy central reasoning and action roles, the relevant behavior is not only an agent's final output, but also the intermediate steps that produce it, from tool calls and data accesses to peer delegations, oversight requests, and graph changes.
For this paper, we use four primary architectural node types as a modeling abstraction for reasoning, action, information access, and human oversight:
\begin{itemize}
\item \textbf{Agents} are LLM-based components that reason or plan over task context and act toward objectives, including by invoking tools~\cite{deng2025agent_security_threats}.
\item \textbf{Tools} are callable functions exposed through structured interfaces, through which agents execute actions including data retrieval~\cite{mcp_architecture}.
\item \textbf{Data nodes} are information sources, persistent or streaming, that agents access for information.
\item \textbf{HITL checkpoints} are runtime review points at which execution pauses to obtain human permission, clarification, or other input before proceeding~\cite{liang2025a2h}.
\end{itemize}

These roles are not an exhaustive ontology of all entities inside an agentic application. Other components such as prompts, skills, and memory may be treated as part of an agent node's internal configuration rather than as separate node types in the application graph. Meanwhile, traditional deterministic software does not disappear in the agent-native paradigm, but rather shifts in its role. Some components become tools that agents invoke on demand, while others may remain as supporting services within the application graph.

The graph is intended to generalize beyond a single agent connected to tools and to account for multi-agent composition as a general case. Multiple collaborating agents can support role specialization, context-window management, and compute efficiency~\cite{kong2025agent_communication, cisco2025internet_agents}. Moreover, recent multi-agent systems show increasing ability to coordinate specialized roles on substantial tasks~\cite{anthropic2026parallel_claudes_c_compiler, cursor2026scaling_agents}, making multi-agent composition increasingly plausible for agent-native application architectures.

\subsubsection{Declarative Application Specification}
\label{sec:declarative-spec}

In cloud-native computing, Kubernetes manifests and service mesh policies declaratively define a desired state, including the services, their scaling and resource bounds, and inter-service communication rules. The platform continuously reconciles the running system toward that state. Agent-native applications similarly benefit from a declarative application specification, but the specification must describe permitted behavior rather than only desired state. It is a specific artifact that defines the application's structure together with the constraints that bound node behavior and graph evolution during execution. We use ``declarative'' in the sense that the artifact specifies these constraints while leaving scheduling, enforcement, and execution strategies to a conforming runtime.

The application specification encompasses two primary concerns. First, the \textbf{application structure} identifies the agents, tools, data sources, HITL checkpoints, and supporting services that constitute the application. It also identifies the nodes that serve as entry points for external requests and the interaction edges among nodes, including the interfaces or communication mechanisms used for those interactions. Second, the \textbf{behavioral envelope} specifies the constraints that govern how runtime agentic reasoning may be exercised and how the application may behave and evolve along several dimensions:
\begin{itemize}
\item Which tools, data sources, peer agents, and external endpoints each agent may access or communicate with.
\item The decision-making autonomy granted to each agent, ranging from tightly constrained workflow sequences to broader runtime discretion.
\item Where human-in-the-loop oversight is mandatory, where it can be dynamically requested by agentic reasoning, and where it is not required.
\item Whether agents may delegate work by spawning sub-agents, and if so, the permissible sub-agent types, maximum number, recursive delegation depth, and permission scope assigned to sub-agents.
\item Which graph changes the application orchestrator may perform during execution, including adding, removing, replacing, or rerouting nodes within declared bounds.
\item Where relevant, resource and execution limits such as cost budgets, compute budgets, maximum execution time, and context or memory bounds.
\end{itemize}

The behavioral envelope is primarily concerned with preventing impermissible proposed actions and traces from becoming realized application behavior, particularly where their realization could produce unsafe or otherwise detrimental outcomes. A trace that remains within the permitted space may still fail to achieve the application's objective with the intended accuracy, quality, or completeness. Evaluating these positive task outcomes requires application-specific success criteria and can be supported in part by the behavioral monitoring and audit mechanisms discussed in Section~\ref{sec:observability}, but it is not the primary concern of the declarative application specification developed here.

\begin{figure}[t]
\centering
\includegraphics[width=\columnwidth]{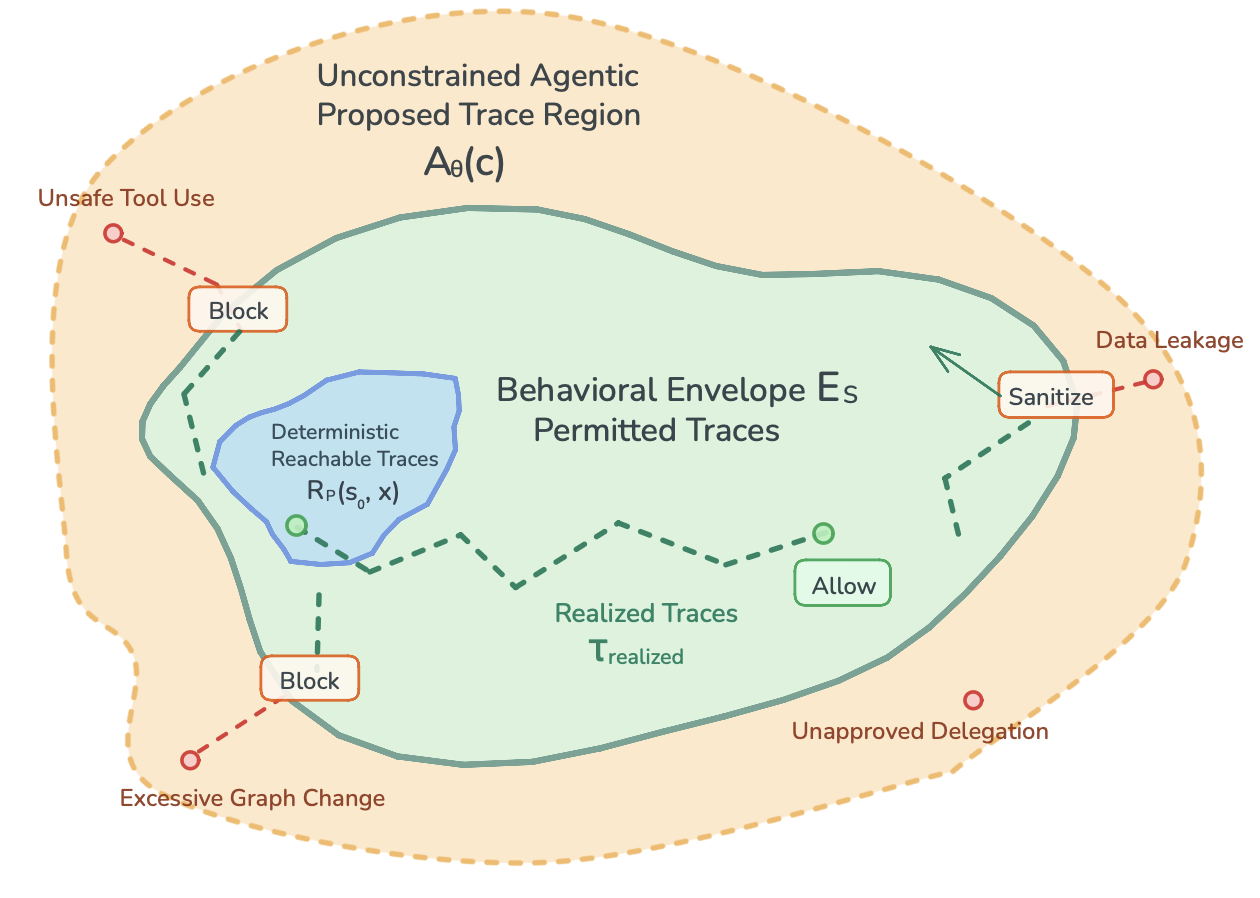}
\caption{Conceptual trace-space view of the behavioral envelope. Deterministic software induces a comparatively small reachable trace set \(\mathcal{R}_{P}(s_0,x)\), while an unconstrained agentic application induces a much larger proposed trace region \(\mathcal{A}_{\theta}(c)\). The behavioral envelope \(\mathcal{E}_{S}\) defines the permitted region of realized behavior, while mediation outcomes such as block or sanitize prevent envelope-violating actions, such as unsafe tool use or data leakage, from becoming realized behavior. The dashed path depicts a single trace unfolding step by step; formally, each complete trace is one element of \(\mathcal{T}\).}
\label{fig:trace-space}
\end{figure}

The specification therefore defines not only the application's initial structure, but also the bounded limits of permissible behavior as the graph changes over time. Because these constraints apply to runtime behavior rather than only to static structure, we can describe the model in trace-space terms. This formalization is intended to characterize which application trajectories may become realized behavior, as illustrated in Fig.~\ref{fig:trace-space}, rather than to provide a complete formal semantics for agent-native execution. Let \(\mathcal{T}\) denote an ambient space of application traces, where an individual trace \(\tau \in \mathcal{T}\) records graph states, agent actions, and HITL checkpoint outcomes over time. The set \(\mathcal{T}\) serves as a conceptual universe for comparing deterministic reachable traces, agent-proposed traces, and envelope-permitted traces.

In the case of traditional software, a deterministic application or component \(P\), initial state \(s_0\), and input sequence \(x\) induce a reachable trace set \(\mathcal{R}_{P}(s_0,x) \subseteq \mathcal{T}\) through the transition relation encoded in code. In contrast, an unconstrained agentic application greatly expands the behavior space because runtime reasoning and actions of agents produce traces that are not enumerated as fixed paths at design time. This space expansion is the source of the application's flexibility, but also creates a major problem when useful, unsafe, or out-of-purpose traces may all be proposed from the same context-dependent reasoning process.

We can view an unconstrained agentic application as inducing a distribution \(p_{\theta}(\tau \mid c)\) over traces, where \(\theta\) abstracts the model and runtime factors shaping behavior, and \(c\) denotes the current prompt, tool, memory, data, and interaction context. We use \(\mathcal{A}_{\theta}(c)\) informally to denote the expanded region of traces that such an application may propose with non-negligible probability under context \(c\). The purpose of the behavioral envelope is to preserve useful latitude within this expanded region while preventing impermissible proposed traces from becoming realized application behavior. For an application specification \(S\), the behavioral envelope \(\mathcal{E}_{S}\) can be expressed as the permitted subset of traces:
\[
  \mathcal{E}_{S} = \{\tau \in \mathcal{T} : \operatorname{Permitted}_{S}(\tau)=1\}.
\]
Operationally, a conforming runtime constrains the agentic application's proposed traces so that realized traces fall within the permitted portion of the expanded agentic region:
\[
  \tau_{\mathrm{realized}} \in \mathcal{A}_{\theta}(c) \cap \mathcal{E}_{S}.
\]
For constraints that the runtime can evaluate reliably, a conforming implementation would enforce this boundary by blocking, sanitizing, escalating, or otherwise preventing attempted actions whose realization would move the application outside \(\mathcal{E}_{S}\).

This trace-space view restates the architectural problem in application-model terms. Agentic execution can propose trajectories that are useful, unsafe, or outside the application's intended scope from the same context-dependent reasoning process. The specification must therefore separate what the application may realize from how the running graph happens to evolve during execution. This separation yields three architectural consequences that support portability across runtimes and frameworks:
\begin{itemize}
\item \textbf{Behavioral envelope versus running graph.} The specification declares what is permitted, while the running graph may change within those declared bounds.
\item \textbf{Portable declarative artifact.} The specification does not prescribe which specific runtime, programming language, or agent framework realizes it. A conforming runtime can consume the artifact and instantiate the application.
\item \textbf{Shared enforcement substrate.} Permissions, human oversight, audit, and graph-evolution constraints are enforced by runtime infrastructure rather than by each node's local implementation alone.
\end{itemize}

These separations keep the agent-native application from being tied to a particular orchestration framework. Meanwhile, agentic nodes built with existing agent frameworks can still participate if they conform to the required interfaces.

\subsection{Runtime Infrastructure}
\label{sec:runtime-infrastructure}

The previous section defined an agent-native application in static terms involving its compositional graph structure and declarative application specification. At runtime, a conforming implementation must instantiate or bind the declared graph with its mediated interaction interfaces, and contain application behavior within the behavioral envelope as tasks proceed.

We divide this runtime responsibility into two complementary planes. The agent mesh is the data plane for inline mediation, while the application orchestrator is the control plane for graph-level coordination, permission adjustment, recovery, and evolution. The main distinction is decision scope within the behavioral envelope. The mesh handles decisions that must occur on the interaction path between application nodes and can be made from the current proposed interaction plus bounded local runtime state. The orchestrator handles decisions that require cross-node history, aggregate behavior, or changes to the graph itself. Fig.~\ref{fig:application-architecture} shows this runtime split within the larger application anatomy, with the orchestrator above the data-plane graph and the mesh positioned on governed interaction paths.

This split is analogous to the separation between control and data planes in cloud-native service mesh architectures~\cite{li2019service_mesh}, but operates on the semantics of interactions rather than on communication mechanics. In a service mesh, data-plane proxies mediate service-to-service calls while a control plane distributes routing and policy configuration to those proxies. The agent mesh instead mediates policy- and audit-relevant semantic interactions, including tool invocations, agent-to-agent messages, and HITL checkpoint interception. For agent-native applications, the application orchestrator remains separate from the agent mesh and functions at a different plane, because its decisions require graph-level state and longer-horizon history, and because it holds the authority to alter the running application graph.

\subsubsection{Data Plane: Agent Mesh}
\label{sec:agent-mesh}

The agent mesh is the data-plane component responsible for inline mediation of policy- or audit-relevant interactions, enforcing the interaction-path portion of the behavioral envelope. The agent mesh has four operational responsibilities:
\begin{itemize}
\item \textbf{Interaction-path policy enforcement.} For interactions routed through the mesh, including tool invocations, agent-to-agent communication, and data access where policy or audit obligations attach, it evaluates whether the interaction is permitted under the declarative application specification. This includes simple access-control checks, such as whether one agent may call another, and context-aware policy evaluation, such as whether an agent may invoke a tool with particular parameters under its current permission scope. The mesh may also use bounded runtime state available on the interaction path, such as whether a prior human approval already covers the proposed action, how many nested delegations have occurred, or how much of a tool-call or cost budget has been consumed.
\item \textbf{Boundary mediation.} When messages cross the application boundary, the mesh applies two directional classes of enforcement. \emph{Ingress enforcement} inspects external-to-application traffic, while \emph{egress enforcement} inspects application-to-external traffic. Section~\ref{sec:boundary-mesh} develops both in more detail as the security and governance realization of this mediation role.
\item \textbf{HITL checkpoint enforcement.} When the specification declares that certain interactions require human approval, the mesh intercepts those interactions deterministically, pauses execution, surfaces the proposed action and relevant context for human review, and resumes execution according to the human decision.
\item \textbf{Event production.} Each mediated interaction produces a structured event carrying information such as action type, agents involved, parameters, enforcement decision, policy evaluation results, latency, and audit lineage. The mesh emits a canonical event stream consumed by the application orchestrator and durable audit storage.
\end{itemize}

Each enforcement decision of the agent mesh falls into one of four outcomes. An \emph{allow} decision lets the interaction proceed unchanged. A \emph{block} decision terminates the interaction and returns a structured rejection. A \emph{sanitize} decision permits the interaction after content modification, such as redaction or parameter clamping. An \emph{escalate-to-human} decision pauses the interaction and triggers human review. This four-outcome model generalizes a binary distinction between allowing and blocking interactions and is based on the taxonomy of allow, deny, sanitize, and escalate decisions~\cite{jackson2025policy_engine}.

The mesh is not a substitute for well-calibrated node behavior. Agentic nodes should be configured locally so that they usually act within their intended roles. Prompts and tool descriptions shape each node's behavior, scoped credentials limit its reach, and node-level guardrails can add local checks. An application in which agents routinely attempt disallowed actions and over-rely on mesh rejection for control would be inefficient and potentially difficult to operate. Meanwhile, the mesh is the shared runtime layer that enforces the interaction-path portion of the behavioral envelope even when node-level behavior fails, drifts, or is compromised. It produces mediated events for audit and orchestration, and handles HITL interception as a runtime concern rather than leaving every node to implement these obligations locally. For mesh enforcement to be comprehensive, all policy-relevant access to tools, data, peer agents, and external services must pass through the mesh.

Mesh policy evaluation can combine deterministic, machine learning (ML)-learned, and LLM-based techniques. Deterministic checks are appropriate for permissions, resource limits, and audit obligations. ML-learned classifiers or content-inspection models can help with semantic risks such as prompt injection or sensitive data leakage. LLM-based reasoning can support novel or ambiguous cases, but does so at higher latency, cost, and stochasticity. The latter two non-deterministic mechanisms should be understood as probabilistic policy enforcement rather than absolute assurance. When sufficiently reliable for a given policy and context, however, they can still provide meaningful runtime enforcement depending on the use case.

\subsubsection{Control Plane: Application Orchestrator}
\label{sec:orchestrator}

The application orchestrator governs graph-level coordination and adaptation within the behavioral envelope. It receives external inputs through designated entry points, including human-provided task objectives or intents, events from external sources, and inter-application requests. These inputs provide the task-level objectives or operating intent that the application works to satisfy. The orchestrator also consumes internal runtime signals, including structured events from the agent mesh, observability telemetry from the application graph and supporting runtime, and capability or permission requests from agentic nodes.

Using this graph-level view, the orchestrator translates objectives, runtime signals, and node requests into control decisions. These include when to reroute work, recover from failure, adjust node permissions, alter graph structure, grant or deny node capability requests, or escalate graph-level decisions for human input. Its decision context extends across tasks and sessions, supporting interventions that cannot be made from a single interaction path. Through these control decisions, the orchestrator keeps graph-level coordination and adaptation directed toward the application's objective or operating intent.

The orchestrator operates within the behavioral envelope declared in the active application specification. It may propose graph changes, capability acquisitions, and permission adjustments as application needs evolve, but those proposals do not take effect on the orchestrator's authority alone. A deterministic reference monitor validates each request against the application specification before the runtime applies it. This separation places the root of trust outside the orchestrator and prevents it from expanding the envelope, disabling mediation, or granting itself additional authority. A change to the behavioral envelope instead requires an update to the application specification that is authorized independently of the orchestrator.

The orchestrator's implementation design choice involves a tradeoff between capability and adaptability on one hand, and reliability and determinism on the other. Three broad forms of orchestrator are possible along this continuum:
\begin{itemize}
\item \textbf{Deterministic.} Pre-programmed workflow logic, with adaptability limited to threshold-based triggers such as scaling the number of active agents under load.
\item \textbf{ML-Learned.} A policy learned from data through an ML model, which adapts based on patterns observed during training. This can include the use of reinforcement learning to orchestrate agents via an evolving policy~\cite{dang2025puppeteer}.
\item \textbf{Agentic.} An LLM-based agentic orchestrator decomposes tasks, selects which agents and tools to engage, spawns new agents, and reroutes task flows based on its own reasoning, enabling the application to address situations not anticipated at design time.
\end{itemize}

An LLM-based agentic orchestrator provides broader adaptability than the other two forms, which react only to conditions anticipated by their designers or training distributions. However, an agentic orchestrator can also introduce reliability issues because stochastic reasoning may produce incorrect judgments, while also incurring higher cost and latency. Given this tradeoff, hybrid designs may be desirable, with deterministic or ML-learned logic handling well-understood workflows and an agentic component handling novel or complex scenarios. At the agentic extreme, if agents become sufficiently reliable, a highly autonomous agentic orchestrator could compose a new task-specific application graph for each incoming request, resulting in a customized application design for handling each task.

The orchestrator's role also resembles the historical MAPE-K loop of autonomic computing~\cite{kephart2003autonomic}, but applied to a qualitatively different kind of managed system. The managed system is an agent-native application graph whose nodes may reason, request permissions, delegate work, and dynamically select execution paths. The orchestrator therefore monitors mesh events, telemetry, and node requests; analyzes and plans whether graph-level intervention is needed; executes graph-level changes within the permitted envelope; and maintains knowledge about the graph model, runtime state, historical observations, and orchestration rules.

\subsection{Connectivity}
\label{sec:connectivity}

\begin{figure*}[t]
\centering
\includegraphics[width=\textwidth,height=0.4\textheight,keepaspectratio]{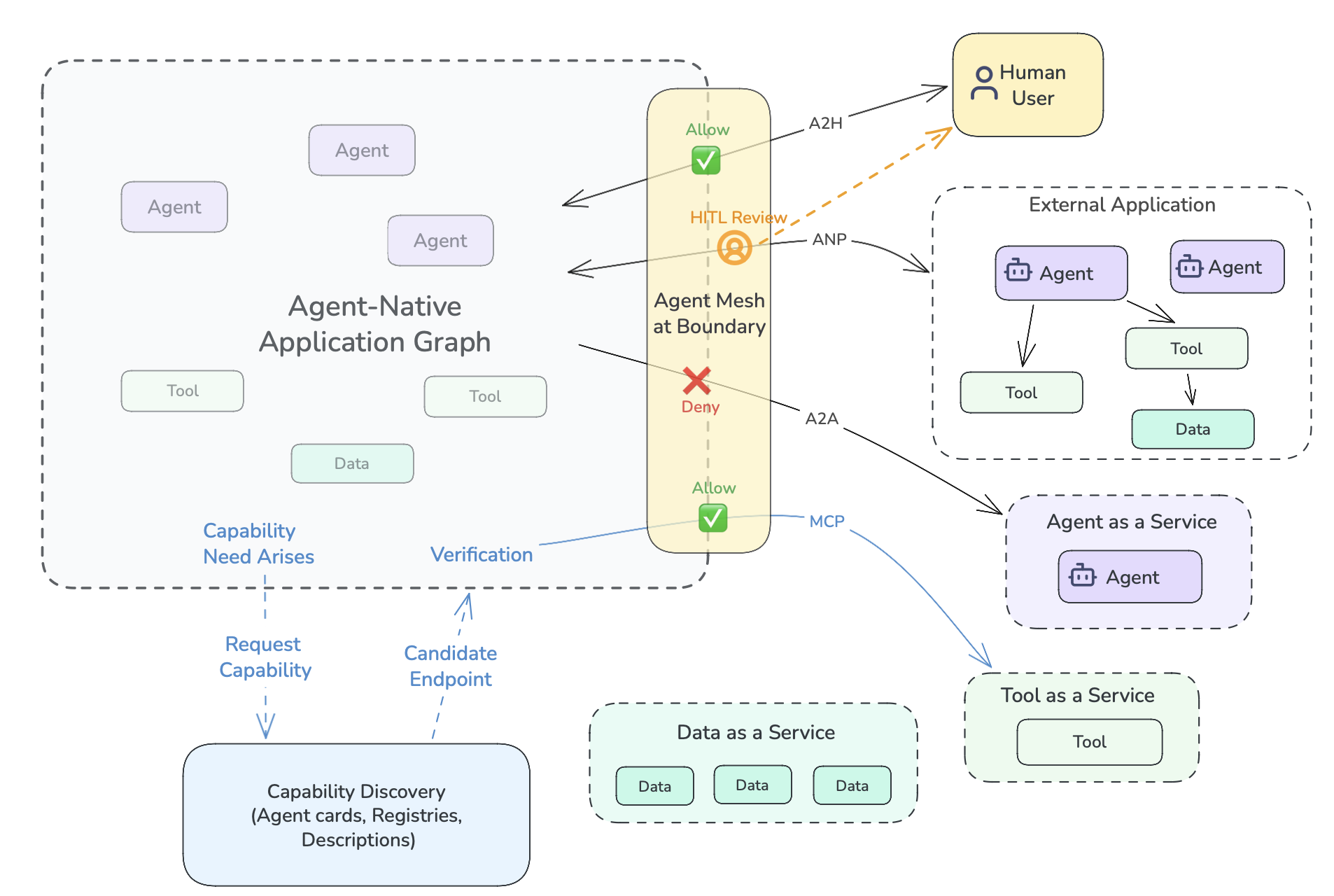}
\caption{Connectivity for agent-native applications. Agent mesh at the application boundary mediates allowed, denied, and HITL-reviewed exchanges with humans, external applications, and agent/tool/data services. Capability discovery identifies candidate endpoints, which are verified before binding.}
\label{fig:connectivity-ecosystem}
\end{figure*}

Agent-native applications require communication among heterogeneous participants, including agents, tools, data services, humans, and external applications. Standardized protocols can reduce fragmentation in these interactions and support collaboration among agents from different providers or architectures~\cite{yang2025survey_agent_protocols, kong2025agent_communication}. MCP, for example, supports structured agent-to-tool invocation~\cite{mcp_architecture}, while A2A~\cite{a2a_protocol_spec} and the Agent Network Protocol (ANP)~\cite{chang2025agentnetwork} illustrate approaches to cross-application agent discovery and task exchange. The agent-native model should therefore interoperate with protocols that provide suitable communication, discovery, identity, or task-exchange mechanisms, without binding the architecture to any one protocol.

The connectivity design should keep the agent mesh loosely coupled from the protocols used by individual nodes. Fig.~\ref{fig:connectivity-ecosystem} depicts this separation between protocol-specific communication and protocol-independent mediation. Protocol adapters, runtime hooks, or gateways can translate particular communication mechanisms into a common mediation interface for interactions subject to policy or audit. In this model, protocols remain interchangeable, while governed interactions are represented to the runtime in a form the mesh can mediate. The mesh remains responsible for enforcement, HITL interception, application boundary inspections, and audit production regardless of which protocol carries the underlying interaction.

The figure also highlights the need to discover external applications or resources at runtime. Some external interactions are address-based because the target endpoint is already known. Others are capability-based, with the application orchestrator searching for an external tool, data source, or peer agent that satisfies a runtime need~\cite{cisco2025internet_agents, guo2026agent_discovery_ioa}. When newly discovered external applications or resources become candidates for task delegation or invocation, the orchestrator must first verify and bind them within the bounds of the behavioral envelope, while the mesh mediates subsequent interactions with them. By analogy to software as a service (SaaS) and related cloud service models, such discoverable capabilities may be exposed as agents as a service, tools as a service, or data as a service.

\subsection{Companion Prototype: ANR}
\label{sec:anr-prototype}

Agent-Native Runtime (ANR) is a lightweight companion artifact that implements a focused subset of the preceding architectural concepts in code. It is not intended as a production runtime or proposed standard. Its role is to demonstrate that a declarative application specification can be compiled into a running agent graph in which runtime infrastructure, rather than node-level prompts alone, mediates tool invocations and HITL checkpoints, enforces scoped authority, resource limits, and graph-evolution bounds, and produces audit records. The ANR prototype is available at \url{https://github.com/agent-native-apps/ANR}.

ANR loads a YAML application specification into a versioned specification model, with a committed JSON Schema serving as the portable contract for the artifact. At compile time, it constructs the declared graph, discovers tool schemas, and derives the runtime authority assigned to each node. At execution time, LLM-based agents and deterministic script-based nodes interact through the same mesh interface. The agent mesh enforces permission scope, HITL checkpoints, boundary policies, and resource limits before dispatching tool calls or accepting graph-evolution requests.

Listing~\ref{lst:spec-fragment} shows an abridged ANR specification fragment, illustrating the central pattern in which scoped node authority is declared as part of the graph, while the envelope defines the oversight, budget, and graph-evolution bounds.

\begin{lstlisting}[
  style=yamlfragment,
  float,
  floatplacement=!t,
  caption={Illustrative fragment of an agent-native application specification.},
  label={lst:spec-fragment}
]
agents:
  - name: coordinator
    tools: [list_docs, grep_files, read_file]
    may_delegate_to: [researcher, writer]
  - name: researcher
    tools: [grep_files, read_file, fetch_url]
  - name: writer
    tools: [read_file, write_note]

envelope:
  autonomy:
    coordinator:
      max_sub_agents: 3
      max_delegation_depth: 2
      max_tool_calls: 25
  resource_limits:
    total_cost_usd: 1.00
    total_runtime_sec: 180
    total_llm_calls: 40
  hitl_checkpoints:
    - id: approve_note_write
      pattern: predefined
      when:
        tool: write_note
    - id: agent_requested_review
      pattern: agent_initiated
      allowed_for: [coordinator, researcher, writer]
  graph_reshape:
    allowed: true
    permitted_changes:
      - kind: spawn_instance
        of_agent: researcher
        max_concurrent: 3
        for_callers: [coordinator]
      - kind: acquire_capability
        capability_kind: tool
        capability_name: fetch_url
        for_callers: [researcher]
        requires_hitl: true
        reason_required: true
\end{lstlisting}

This implementation realizes the three separations developed in Section~\ref{sec:declarative-spec}. The specification remains an application-level artifact, while node implementations can be LLM-based agents, deterministic script-based nodes, or other node types conforming to the same interface. Tools may be exposed through MCP or simple Hypertext Transfer Protocol (HTTP) bindings. The same decoupling applies at runtime, where node authority is carried in the invocation context rather than in node configuration, graph-evolution requests are checked against the declared rules, and mediated interactions produce structured audit events.

The repository includes executable specifications for several scenarios, including the emergency-response and supply-chain examples discussed in Section~\ref{sec:use-cases}. To illustrate the above mechanisms during execution, we ran repeated executions of the three specifications summarized in Table~\ref{tab:anr-scenarios} and computed descriptive average counts from the resulting audit logs. Claude Haiku 4.5 was used to power the agents and a total of 50 runs were executed for each specification, half of them with fault-injection and half without. Human-review checkpoints used a scripted auto-approve backend, recorded in the audit trail, so that runs could complete without manual intervention. The scripted auto-approve backend demonstrates checkpoint interception, decision routing, and audit recording, but does not evaluate realistic human judgment or response latency. Across the runs, different parts of the behavioral envelope were exercised. Emergency response instantiated the hazmat coordinator through a declared graph-evolution rule, supply chain triggered ingress and egress boundary mediation around supplier content, and fault-injected runs exposed broad tool schemas while leaving mesh authority scoped, producing refused out-of-scope action proposals that would otherwise have executed. The full audit logs and more details are available in the provided repository. The table reports these audit-derived counts to characterize the prototype's runtime behavior. They are not a performance or security evaluation.

\begin{table}[t]
\caption{ANR scenario specifications and audit-derived runtime activity across unattended runs. Per-run values are averages. HITL decisions used the scripted auto-approve backend and are recorded as such in the audit logs. The fault-injection row reports 25 runs per scenario in which all declared tool schemas were exposed to every agent while mesh-side authority remained scoped; $x \rightarrow y$ denotes proposed out-of-bounds actions, with $x$ proposed by agents and $y$ realized after mesh mediation.}
\label{tab:anr-scenarios}
\centering
\footnotesize
\setlength{\tabcolsep}{1pt}
\begin{tabular}{lccc}
\hline
 & \shortstack{research\\assistant} & \shortstack{emergency\\response} & \shortstack{supply\\chain} \\
\hline
Nodes (agents/tools/data) & 3 / 5 / 2 & 5 / 7 / 2 & 5 / 8 / 3 \\
Envelope rules (HITL/reshape/budget) & 3 / 4 / 3 & 5 / 3 / 3 & 5 / 2 / 3 \\
Mesh-mediated tool calls per run & 16.8 & 26.0 & 32.2 \\
HITL episodes per run & 1.3 & 4.3 & 9.3 \\
Graph-reshape ops per run & 0 & 1.0 & 0.5 \\
Boundary decisions (allow/sanitize/block) & -- & -- & 83 / 45 / 0 \\
Agent actions refused & 2 $\rightarrow$ 0 & 7 $\rightarrow$ 0 & 7 $\rightarrow$ 0 \\
\quad same, fault-injected runs & 56 $\rightarrow$ 0 & 45 $\rightarrow$ 0 & 86 $\rightarrow$ 0 \\
Distinct traces / runs & 22 / 25 & 25 / 25 & 25 / 25 \\
\hline
\end{tabular}
\end{table}

ANR is intentionally scoped as an application-layer feasibility prototype. It runs as a single-process reference implementation with simplified policies and a bounded set of graph-reshape operations. Distributed deployment, production hardening, scalability evaluation, and full cross-application protocols such as A2A or ANP remain for future implementations.

\subsection{Open Questions}
\label{sec:agent-native-open-questions}

The preceding subsections defined the application graph, behavioral envelope, runtime plane split, connectivity model, and ANR prototype. A number of questions remain before this agent-native application layer can be realized as a portable and governable architecture across runtimes.

\textbf{Behavioral envelope semantics and enforceability.} Deterministic envelope constraints, such as fixed tool permissions or numeric budgets, can usually be expressed as explicit predicates once the specification syntax is defined. The harder open question is how to specify and enforce constraints that require contextual judgment, such as whether an outbound message discloses sensitive intent or whether a proposed action remains within the application's purpose. These cases may require ML or LLM-based evaluation, which introduces challenges in defining the constraint semantics, evaluator obligations, and runtime conformance criteria.

\textbf{Functional specification.} Agent-native applications retain functional requirements even when agents select solution procedures at runtime, and behavioral-envelope conformance alone does not establish functional success. An open question is whether and to what extent these requirements should be addressed at the application-architecture level, including whether declarative application specifications should encode objectives and success criteria and how validators, monitors, or HITL checkpoints should participate in assessing them.

\textbf{Applicability boundaries.} A related but distinct question concerns which application functions are well suited to agent-native execution. Functions that benefit from contextual judgment and adaptable execution may be natural candidates, while those with strict requirements for execution sequence, timing, or physical control may remain better suited to deterministic components or hybrid designs.

\textbf{Runtime graph evolution.} Runtime graph evolution provides agent-native applications with adaptivity, but unconstrained evolution could result in the running graph drifting beyond the intended application form. The key open question is how the behavioral envelope should express bounds on this evolution. Possible mechanisms include hard transformation rules, typed agent templates, policy predicates, prompt-based guidance to an agentic orchestrator, or combinations of these approaches. A useful specification should allow graph changes for legitimate runtime needs while making clear the restrictions on new nodes, edges, permissions, and contexts that may be introduced.

\textbf{Mesh realization.} The agent mesh has been proposed as a logical mediation plane, but its physical realization remains open. Some possible runtime options include inline mediation of structured interaction traffic, a broker that invokes an agentic communication protocol after policy evaluation, framework hooks before tool calls or agent-to-agent interactions, and gateway mediation at application boundaries. These options differ in latency, deployment complexity, observability, enforcement coverage, bypass resistance, and compatibility with existing frameworks. A useful standard should therefore specify the mediation contract and enforcement obligations without prematurely fixing on a specific type of physical realization.

\textbf{HITL checkpoint protocol.} The agent mesh can determine when human review is required. However, we are currently not aware of a commonly adopted checkpoint protocol for the HITL interaction itself. Such a protocol would define how execution pauses, what context is shown to the reviewer, how the decision is represented, and how execution resumes. The Agent-to-Human (A2H) protocol~\cite{liang2025a2h} is one proposal in this direction, but these efforts remain at an early stage.

\textbf{Capability discovery, verification, and binding.} Capability-based application or resource discovery can be a powerful source of runtime flexibility when the application has a capability need that was not provided for at deployment time. An area to address is how the envelope should specify which discovered capabilities are acceptable, how candidates should be verified, and how selection should occur when multiple candidates satisfy a need. After selection, the runtime must also decide what permission scope the capability receives, whether it remains an external invocation or becomes part of the graph, and how subsequent interactions are mediated and audited.

\section{Trust Layer: Security, Governance, and Observability}
\label{sec:trust-layer}

\begin{figure*}[t]
\centering
\includegraphics[width=\textwidth,height=0.48\textheight,keepaspectratio]{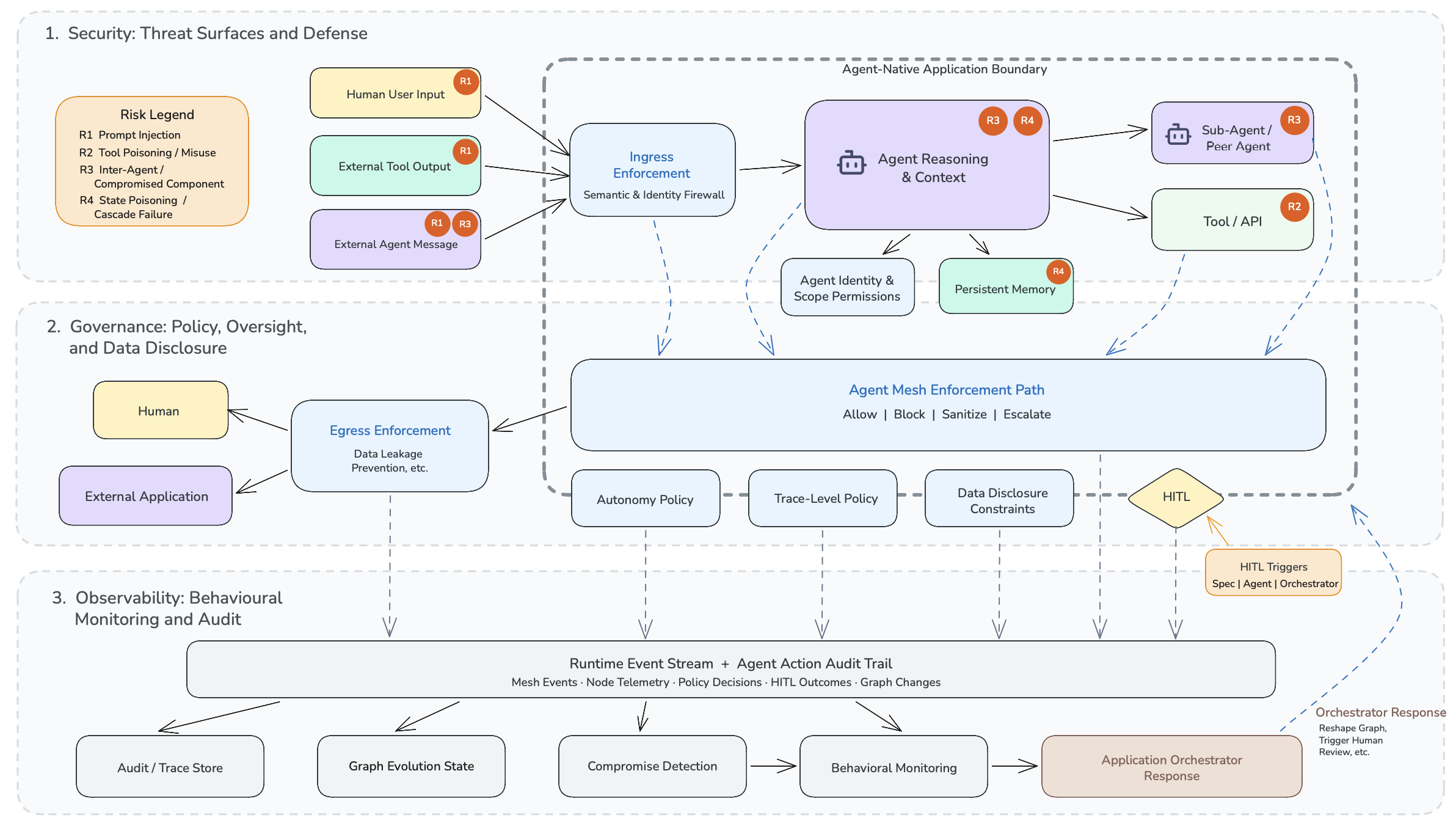}
\caption{Trust layer mechanisms in an agent-native application. Security maps external inputs and agent-specific threats, along with defenses including ingress enforcement and agent identity. Governance applies agent mesh enforcement, egress controls, policies, and HITL oversight. Observability turns runtime events and traces into audit, monitoring, and response signals.}
\label{fig:trust-mechanisms}
\end{figure*}

On the basis of the agent-native application layer, the trust layer focuses on keeping the application operationally dependable. Its concerns span security, governance, and observability, as mapped in Fig.~\ref{fig:trust-mechanisms}. Security addresses agent-specific threat surfaces and defense mechanisms. Governance addresses autonomy, human oversight, and data disclosure. Observability addresses agent behavioral monitoring and audit.

\subsection{Security}
\label{sec:security}

\subsubsection{Agentic Threat Surfaces}

With agentic nodes as the centerpiece of agent-native applications, new security threats arise from LLM agents reasoning over context and acting through tool calls and peer messages. Security must therefore defend each path through which malicious content or compromised state can influence application behavior. While conventional cloud security measures such as network protection and dependency management remain necessary, here we focus on novel security needs that arise in agentic applications.

For illustrative purposes, we highlight several representative agentic threat surfaces commonly discussed in existing taxonomies~\cite{owasp_agentic_applications, kong2025agent_communication, deng2025agent_security_threats}:
\begin{itemize}
\item \textbf{Prompt injection.} Because LLM agents condition their behavior on natural-language context, agent reasoning and actions can be hijacked by adversarial instructions embedded in user input, retrieved documents, websites, tool outputs, or messages from peer agents~\cite{greshake2023indirect_prompt_injection}.
\item \textbf{Tool poisoning and misuse.} Malicious or misleading tool descriptions can steer an agent's tool selection or parameters. Compromised tool endpoints can return outputs that manipulate later agent behavior~\cite{shi2026toolhijacker}.
\item \textbf{Inter-agent and compromised-component attacks.} Communications or trust between agents can be exploited by peer-agent messages, newly discovered external agents, or compromised application components.
\item \textbf{State poisoning and cascading failure.} Corruption of memory, persistent state, or prior outputs can propagate across later tasks and agent-to-agent interactions, allowing one compromised interaction to have a cascading impact on graph activities~\cite{owasp_agentic_applications}.
\end{itemize}
Note that these examples are not intended as a comprehensive list of threat categories, which would be far broader and is beyond the scope of this paper.

Defending against agent-specific threats like these is necessary for agentic applications to run safely. The agent-native architecture should address them through layered controls, which the subsequent subsections address. Scoped permissions place restrictions on the actions that a compromised or prompt-injected agent can perform, thereby containing its blast radius. Identity and bounded delegation make agent-to-agent trust and permission propagation explicit rather than implicit. The agent mesh, acting as a firewall, can block incoming communications that appear malicious. Provenance, trace monitoring, and audit trails help detect unusual behavior and support containment when prevention is incomplete. The goal is therefore to reduce the risk of unsafe proposed behavior becoming realized behavior.

\subsubsection{Scoped Permissions and Identity}

Scoped permissions and identity provide an architectural defense against abuse. Each agent is assigned an identity distinct from the user, service, model endpoint, or parent agent that caused it to run~\cite{du2025agent_communication, cao2026auton}. The identity anchors authorization and audit, and provides other agents or applications with a basis for deciding whether a peer should be trusted. In the agent-native architecture, the declarative application specification should assign or configure an identity for each agentic node, together with its associated permission scope. When the orchestrator spawns a new agentic node at runtime, it should likewise provision that node with an identity and bounded permissions appropriate to the delegated task. In networked agent environments, risks at the identity layer, such as impersonation and Sybil attacks, can also arise. Decentralized identifiers and verifiable credentials are possible foundations for tamper-resistant identity management~\cite{wang2025ioa_security}.

Sub-agent delegation is a particular case where permission scoping can be challenging because permissions can be requested and propagated at runtime. A sub-agent should receive only the tools, data, and peer interactions needed for its delegated task, rather than inheriting the full access context of the agent that spawned it~\cite{owasp_agentic_applications}. Since the parent agent may decide the task and required context dynamically, the runtime cannot rely solely on static role assignment. The orchestrator must validate requested permissions against the behavioral envelope and task need, while the mesh enforces the resulting permission scope. This limits the blast radius of prompt injection, compromised peer agents, and excessive delegation by ensuring that a successful influence attempt does not automatically inherit broad authority within the graph.

\subsubsection{Agent Mesh as Firewall}
\label{sec:boundary-mesh}

As mentioned in Section~\ref{sec:agent-mesh}, the agent mesh mediates interactions among application nodes and inspects exchanges that cross the application boundary. Its boundary role is analogous to an agent-facing firewall. Unlike a conventional transport firewall, its focus is on inspecting the meaning and content of incoming communication from external agents or tools, including messages that may contain prompt injection~\cite{fleming2025layered_protocol_iot_agents, chennabasappa2025llamafirewall}. It can also consider agent identity and policy context before allowing a boundary-crossing exchange to proceed~\cite{twingate2025identity_firewall}. This role becomes especially important when external capabilities are discovered at runtime. Candidate agents, tools, or data services must be verified before they are bound into the graph, and later interactions with them must remain subject to mediation.

For security concerns, the main emphasis is ingress enforcement, which inspects external content reaching the application graph. The outgoing disclosure-oriented side of application boundary mediation is treated in the next section mainly as a governance concern.

\subsection{Governance}
\label{sec:governance}

Governance defines what the application is permitted to realize. Even without malicious attack, an agent may exceed its intended scope, take high-stakes action without human approval, or reveal information that should remain confidential. In an agent-native application, governance becomes operational when runtime decisions can be evaluated as policy at mesh mediation points or orchestrator control points. It therefore centers on policy enforcement, human oversight, and data disclosure control.

\subsubsection{Behavioral Envelope as Policy}

Formulating behavioral envelope constraints as policy rather than natural-language guidance allows them to be evaluated and enforced at runtime, while making the resulting decisions auditable. In an agent-native runtime, governance policies cover concerns such as the extent of autonomy granted to agents, the limits on information disclosure to external applications, and when human review is required. These constraints can be expressed through guardrails, agent-specific DSLs, path-based governance functions, or behavioral contracts~\cite{rebedea2023nemo_guardrails, wang2025agentspec, kaptein2026governance, bhardwaj2026agent_behavioral_contracts}. Within the architecture proposed in this paper, the declarative application specification encodes these policies as part of the behavioral envelope, while the agent mesh and orchestrator evaluate proposed actions or graph changes against those policies.

Policy checks on individual actions are necessary, but some governance questions depend on the sequence in which actions occur. For example, a database read may be acceptable in isolation, while the same read followed by an external message may indicate possible data exfiltration~\cite{kaptein2026governance}. Trace-level evaluation is therefore needed when the meaning or risk of an action depends on what preceded it or what it enables next.

Consistent with the runtime split in Section~\ref{sec:runtime-infrastructure}, such evaluation may involve both mesh-level checks on the interaction path and orchestrator-level assessment over longer-running traces. Some evaluations can be hardcoded, but trace-level judgment may involve ambiguous situations that cannot be judged with certainty. Responses should therefore be proportional to the seriousness of the use case and the consequences of an incorrect decision. When the risk is significant or the consequences are difficult to reverse, the mesh or orchestrator should surface the case to a human for direction.

\subsubsection{Human Oversight}
\label{sec:hitl-oversight}

Human oversight is necessary when runtime agentic reasoning needs human authority, contextual judgment, or clarification of intent. In some cases the issue is consequence, such as a high-stakes or irreversible action. In other cases, the agent may need human input to resolve ambiguity or calibrate priorities, so as to avoid spending substantial effort on a misaligned path. The design problem is therefore deciding when execution should pause for review and when it should continue autonomously. Review must be available when human judgment materially affects the safety or usefulness of the outcome, while routine work should proceed autonomously when it remains within the declared envelope. Excessive oversight reduces the efficiency benefit of delegating tasks to LLM agents and can also degrade review quality when approval requests become overly frequent or time-pressured~\cite{kaptein2026governance}. Structured autonomy taxonomies can help define the level of the human's role~\cite{feng2025autonomy}.

Recalling the application-layer model, human oversight is practically represented through checkpoints under the behavioral envelope as well as dynamic review requests. A review episode pauses the relevant action, surfaces context to the reviewer, records the decision, and resumes execution according to the reviewer's input. Human oversight can arise in several ways:
\begin{itemize}
\item \textbf{Specification-defined checkpoints.} The declarative application specification can require review for predefined actions, with the agent mesh enforcing the checkpoint when the condition is reached.
\item \textbf{Agentic-node requests.} An agentic node within the application graph can request human input when it encounters uncertainty or needs clarification of intent. The request should be routed through the orchestrator, which decides whether to escalate it to a human.
\item \textbf{Orchestrator-initiated review.} The orchestrator can initiate manual review when graph-level patterns, such as unusual behavior or accumulating cost, warrant human attention.
\end{itemize}
Together, these sources of review support both predefined review points and dynamic situations that surface at runtime.

\subsubsection{Data Disclosure Boundaries}

Data exiting the application is a major risk point for leakage of confidential or private information. The core governance problem is an asymmetry between access and disclosure, where agents may need internal data to reason effectively, but must not disclose that data to external parties. However, internal context used for reasoning can be carried into outbound messages in several ways. An agent may quote or paraphrase sensitive material retrieved from an internal knowledge base, or it may reveal sensitive facts indirectly through a summary, recommendation, or negotiation message. Leakage can also occur through agents that have been fine-tuned on sensitive data or poorly sanitized datasets~\cite{wang2025ioa_security}. In such cases, controlling access to data sources alone is insufficient because the agent may have a legitimate reason to read the information, or the information may already be stored in the model weights. Governance must therefore inspect and gate data at the exit point of the application.

The application specification needs to encode the disclosure boundary as part of the behavioral envelope. It should distinguish information that may be made public from information that is internal or confidential. For the latter, the specification should indicate whether disclosure is always prohibited or allowed for approved recipients under specified conditions. In practice, the agent mesh enforces this boundary in a governance-focused egress role. The mesh inspects outbound messages before transmission and can redact sensitive content, block the message and ask the sender agent to revise it, or escalate to a human for review. Determining whether a transmission violates confidentiality may require deterministic conditional checks over labels or patterns, combined with agentic reasoning to handle more complex logic.

\subsection{Observability}
\label{sec:observability}

Security and governance define what the runtime should prevent and permit. Observability provides the evidence needed to determine whether those boundaries remain effective, whether agents are performing competently, and when the orchestrator should intervene. While conventional infrastructure observability for uptime, latency, and dependency health remains necessary, this section focuses on requirements specific to agent-native applications, including visibility into mediated actions, delegation lineage, graph evolution, and behavioral quality.

\subsubsection{Behavioral Monitoring and Compromise Detection}

Behavioral monitoring tracks whether agents remain grounded in their responses, choose tools appropriately, and interact with peer agents in ways consistent with their roles. It can also monitor whether agents consume resources within expected bounds and whether they drift over time from their intended role or behavior. Methods such as semantic entropy~\cite{farquhar2024semantic}, cognitive observability~\cite{rombaut2025watson}, and behavioral drift metrics~\cite{bhardwaj2026agent_behavioral_contracts} can provide signals of competence and calibration. They help the orchestrator and human operators assess the current state of the application. When the orchestrator detects situations requiring intervention, it can respond through graph-level actions or by requesting a human decision.

Observability also helps detect when an agentic node may be compromised and supports response by connecting its behavior to action lineage and downstream effects. A compromised agentic node may manifest through anomalous tool call sequences or suspicious messages sent to peer agents, even when it remains available and produces plausible routine outputs. Retrieval-augmented generation (RAG) backdoor attacks are one example, where poisoned retrieved content can lead agents to answer user queries correctly while executing hidden malicious commands in the background~\cite{lupinacci2025inter_agent_trust}. Approaches such as goal-conditioned drift detection can be applied for distinguishing legitimate behavioral adaptation from suspicious shifts~\cite{wang2025mi9}. When risk is detected, the orchestrator can reduce node permissions, request human review, reroute work, or isolate the affected node.

\subsubsection{Audit Trails and Agent-Native Traces}

Agent action audit trails aggregate records from across the runtime, including agent mesh events, orchestrator decisions, graph-evolution records, model and tool invocations, policy decisions, HITL outcomes, and node-local telemetry. These records can support investigations of compromise and cascading failures, verify permission and HITL compliance, account for resource and cost usage, and reconstruct graph evolution across deployments.

Agent-native traces should organize these audit records into causal decision paths. This role is analogous to distributed tracing in cloud-native systems, but the object being reconstructed is a decision path rather than a request path across services. Such a path may connect prompts, retrieved context, tool calls, policy decisions, and delegations that led to an output or action. Because these records can themselves contain sensitive content, trace systems should support redaction, retention limits, or summary-level provenance. Recent work points toward different parts of this representation problem, including structured agent logging~\cite{alsayyad2026agenttrace}, provenance graphs for agentic workflows~\cite{souza2025provagent}, and agent-semantic telemetry for runtime governance~\cite{wang2025mi9}. A common event representation would make such traces usable for audit, investigation, and runtime response across runtimes.

\section{Foundational Layer}
\label{sec:foundational-layer}

Having defined the agent-native application model and trust mechanisms, we now turn to the data, cloud, and networking foundations on which the application runs. While these foundations are not a part of the core agent-native architecture, they can substantially impact the performance and practical operation of the application.

\subsection{Data Foundations}
\label{sec:data}

\begin{figure}[t]
\centering
\includegraphics[width=\columnwidth]{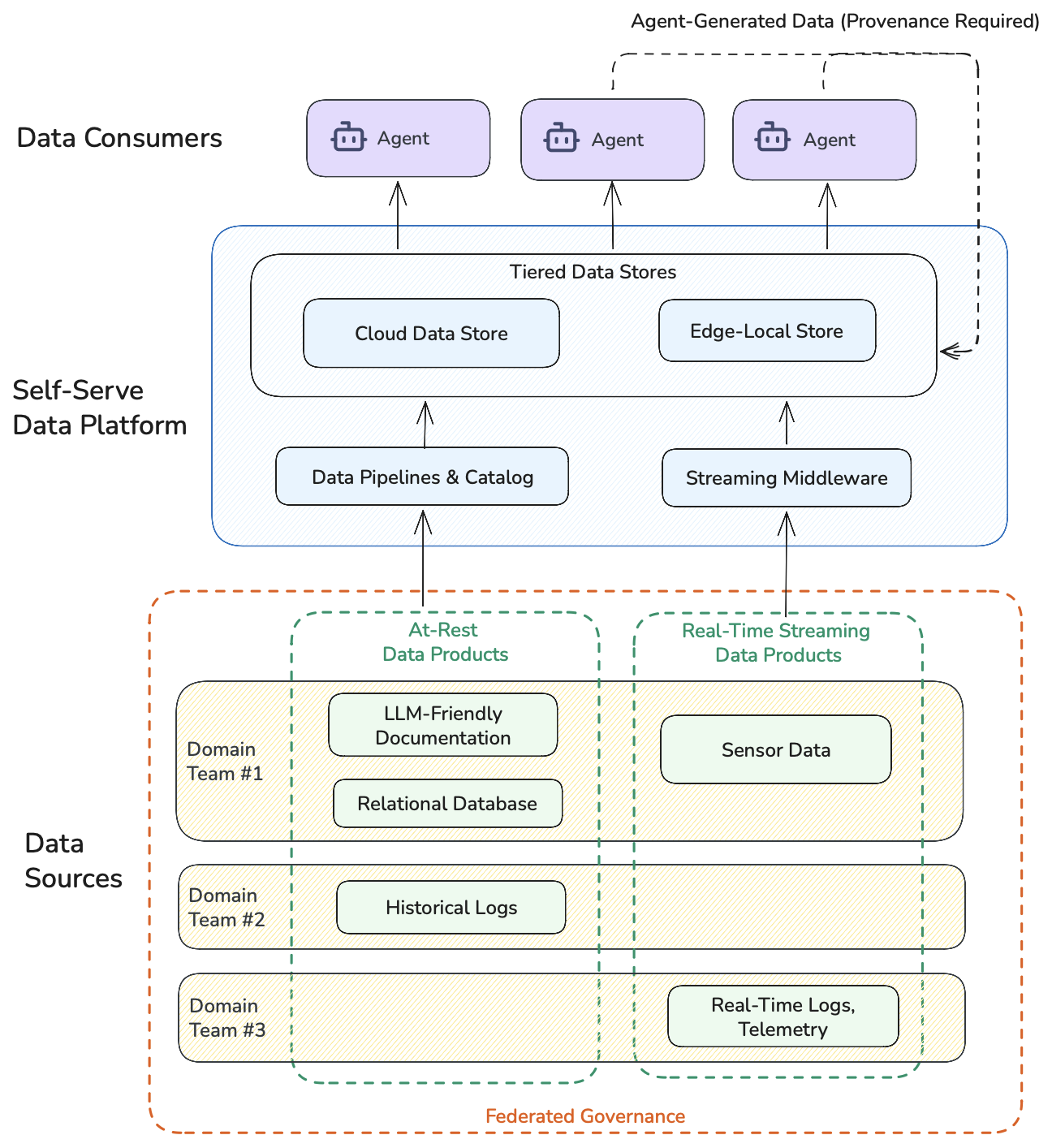}
\caption{Data foundations for agent-native applications. Domain teams publish at-rest and streaming data products under federated governance; a self-serve data platform tiers cloud and edge stores, and preserves provenance for agent-generated data.}
\label{fig:data-foundations}
\end{figure}

For domain-specific tasks, agents often need access to data such as documentation and database contents. Across organizations and institutions, however, these sources may be fragmented, inconsistently formatted, partially outdated, or missing the metadata needed to assess reliability~\cite{bushaus2025modern_data_architecture}. The central issue is therefore not merely whether agents can connect to a data source through a protocol such as MCP, but whether the data itself is effectively organized for agent use. Agents benefit from data environments that support source discovery, semantic interpretation, freshness and authority assessment, sensitivity handling, and provenance tracking across later reasoning and audit. Fig.~\ref{fig:data-foundations} illustrates these data foundations, which are covered in the following subsections. Without these properties, agents may retrieve incomplete evidence or reason from stale records. When data quality is uneven, they can also lack the quality and provenance signals needed to judge which sources to trust.

\subsubsection{Data Mesh and Data Products}

Data mesh and data products provide one systematic approach to the need for agent-accessible data that is discoverable, interpretable, governed, and reusable. Data mesh shifts enterprise data architecture toward domain ownership, where teams closest to the data publish and maintain data assets through shared platform services and federated governance~\cite{goedegebuure2025data_mesh_review, machado2022data_mesh}. Data products are the consumption-facing units of this architecture. They package data and metadata, possibly including access endpoints, quality metrics, lineage information, and access policies, into managed assets such as curated datasets or event feeds~\cite{goedegebuure2025data_mesh_review, machado2022data_mesh, blohm2024data_products_mesh}. For agentic workloads, this packaging is especially useful because agents may need to select sources autonomously, interpret what the data means, and carry quality, sensitivity, and provenance context into later reasoning, generated artifacts, and audit records.

\subsubsection{Real-Time Data Delivery}

Many agentic workloads are operations-oriented, meaning that agents act against external conditions that change over time. Agents monitoring industrial equipment, coordinating emergency response, or overseeing financial systems depend on timely data to assess current state and choose appropriate actions. Streaming middleware such as Kafka and Data Distribution Service (DDS) can be used to provide continuous delivery for such workloads, enabling data freshness for safe and effective task execution.

In edge deployments, routing streaming or frequently updated operational data through centralized cloud storage can add latency and may conflict with data-residency constraints. Edge-local data stores and virtualized edge data layers, such as EdgeLake~\cite{lfedge_edgelake}, allow streaming data to be stored close to the agents and physical processes that depend on it while still being exposed through queryable interfaces.

\subsubsection{Agent-Generated Data and Provenance}

Traditional data frameworks have been largely designed to serve human analysts, deterministic software, and conventional ML pipelines. LLM agents, however, are a new type of data participant because they not only consume data but also produce artifacts that can re-enter the data ecosystem~\cite{equinix2026data_problem_agentic_ai}. Their outputs, including summaries, analytical results, and derived datasets, may be consumed downstream by other agents or humans. Such outputs should be explicitly labeled as agent-produced and accompanied by provenance, reliability signals, and the context used during generation. Sustaining this traceability requires audit trails covering both the data consumed by agents and the data produced by them, so that agent-generated content can be traced back to its inputs and to the conditions under which it was produced.

\subsection{Cloud Deployment and Orchestration}
\label{sec:cloud-deployment}

Once an agent-native application has been defined, it must be deployed onto physical infrastructure. Agent-native deployments could span a cloud-edge continuum from centralized cloud regions to higher-capacity edge clouds and more constrained local edge nodes, driven by application requirements such as latency, cost, data sovereignty, and proximity to users or physical processes~\cite{varghese2017edge_service, zhang2022eaas}. Centralized regions are mature and provide scale, while edge clouds can place computation close to users and data, though with possibly tighter compute and power constraints~\cite{morabito2016edge_iot, varghese2017edge_service}. More specialized settings, including satellite edge~\cite{li2021edge_leo} and vehicle edge computing~\cite{he20216g_cav}, may extend the continuum into non-terrestrial and vehicular environments in the future.

Cloud deployment foundations therefore mainly concern the placement of the application's nodes and supporting services across this cloud continuum. At resource-constrained edge tiers, local capacity may be insufficient to run larger LLMs. Distributed LLM inference techniques such as split inference and collaborative edge inference can enable running such models at the required edge tier~\cite{xue2024wireless_moe, xu2024llm_agents_6g}. Deployments may also span multiple cloud providers, motivated by accelerator availability, cost, regulatory requirements, and the limited geographic reach of any single provider's edge presence~\cite{alonso2023multi_cloud_native}.

Geographically distributed deployment also raises the question of where the runtime infrastructure itself is placed. Because the agent mesh mediates interactions inline, its enforcement points benefit from placement near the nodes whose interactions they mediate, analogous to service mesh data planes. The application orchestrator presents a more difficult placement problem because it holds graph-level authority and state. Operating the orchestrator as a single instance at one site can position it far from graph nodes in other regions, adding network latency to control decisions and making the application dependent on connectivity to that site. Cross-site deployments may therefore require distributed or hierarchical realizations of the orchestrator and mesh, which must in turn maintain a consistent view of shared envelope state, such as application-wide cost budgets, across enforcement points. We identify this as a deployment consideration for conforming runtimes rather than prescribing a specific design.

\subsection{Networking for Agentic Applications}
\label{sec:networking}

Networking demands can arise within an application graph, whose nodes may be deployed across regions, edge sites, or cloud providers. They also apply to messaging across application boundaries, where agents communicate with external agents, tools, and data services.

Latency is a primary concern because a task may be completed through many agentic steps, meaning the network delay of each step accumulates across the entire execution. Meanwhile, agents handling streaming multimodal inputs and outputs, such as audio and video, remain sensitive to delay. Traffic classification and quality-of-service (QoS) mechanisms may therefore need to recognize agentic flows as distinct from undifferentiated best-effort traffic, with software-defined networking (SDN) and network function virtualization (NFV) providing the programmable substrate for such treatment~\cite{bonfim2018integrated_nfv_sdn}. Agents requiring tight latency guarantees, such as those involved in physical control, can employ bounded-latency mechanisms such as Time-Sensitive Networking (TSN) and Ultra-Reliable Low-Latency Communication (URLLC)~\cite{finn2018tsn, zanbouri2025wireless_tsn}. Path control becomes relevant when a data-sensitive agent invocation needs assurance about the jurisdictions or transit domains crossed en route, and Segment Routing over IPv6 (SRv6) can help express such path objectives within cooperating network domains~\cite{ventre2021segment_routing}. Lastly, deployments in disrupted or remote environments, such as disaster zones and remote industrial sites, can rely on satellite-terrestrial networks for resilient network reach when terrestrial connectivity is degraded~\cite{zhu2022sat_terrestrial_6g}.

\section{Use Cases}
\label{sec:use-cases}

This section illustrates the contents of the preceding sections through two use-case scenarios in which agents act as central reasoning and coordination components. Emergency first response exercises the infrastructure and networking demands of edge-deployed environments while also requiring time-sensitive, high-stakes human oversight. Cross-organizational supply chain procurement provides a contrasting enterprise setting where the central concerns are governance, confidentiality, and boundary-mediated negotiation. Together, the scenarios illustrate how the application, trust, and foundational layers converge in agent-native applications with differing requirements.

These scenarios illustrate how the proposed agent-native application architecture can organize and govern agentic reasoning in emergency-response and supply-chain settings. Deployment readiness depends on factors beyond the architecture itself, including the maturity and capabilities of the underlying LLM models, the reliability of supporting data, infrastructure, and control systems, and the training and operating procedures of human reviewers who collaborate with agents. Within such deployments, functions requiring exact behavior, predictable timing, or precise physical control may be more appropriately implemented as deterministic software components.

The companion ANR prototype includes executable YAML specifications for both scenarios, providing inspectable reference artifacts for the declarative application specification, behavioral-envelope constraints, and runtime-evolution semantics. Table~\ref{tab:anr-scenarios} in Section~\ref{sec:anr-prototype} reports measured runtime activity for both specifications. Implementation details and instructions for execution are provided in the repository.

\subsection{Emergency First Response}

\begin{figure*}[t]
\centering
\includegraphics[width=\textwidth]{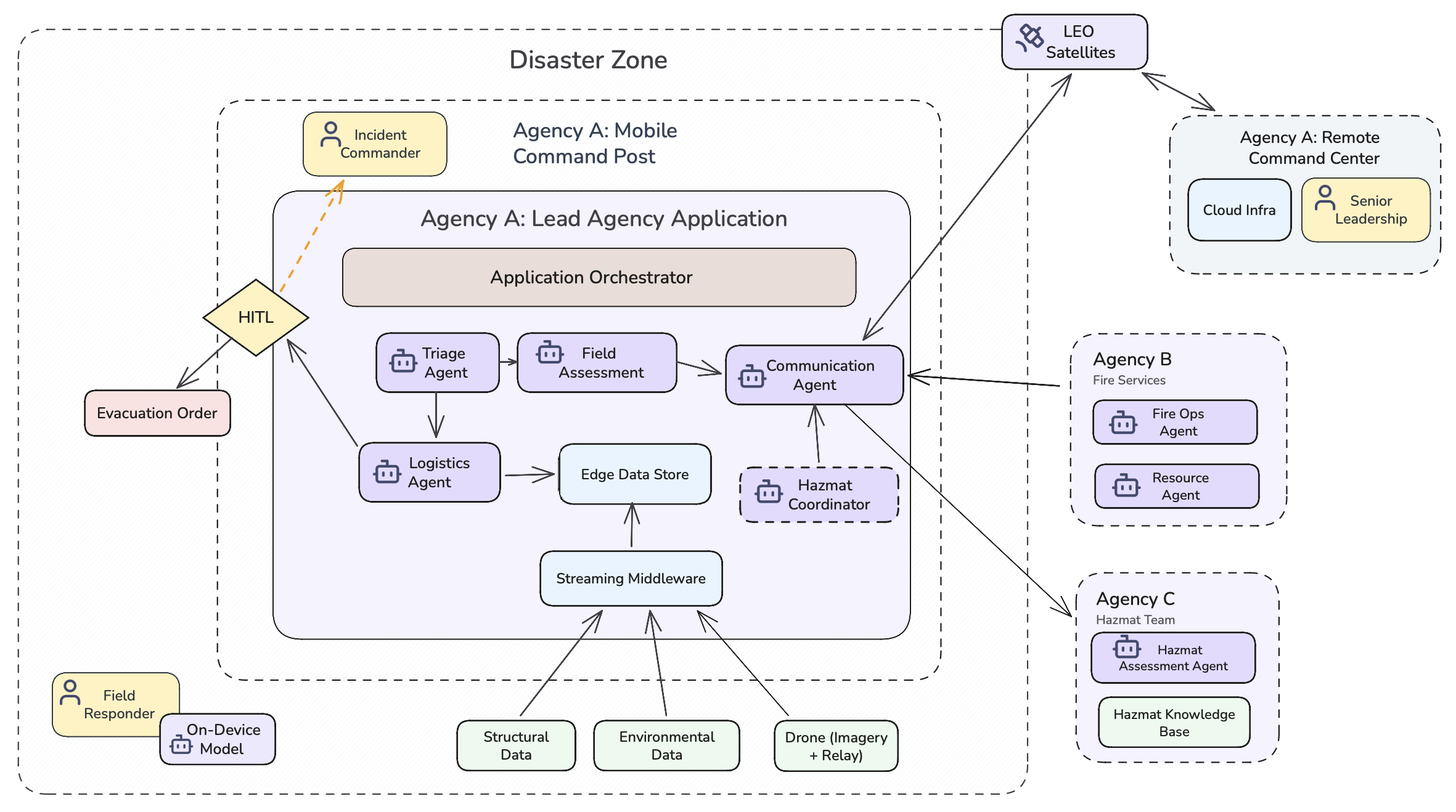}
\caption{Emergency first response deployment for an agent-native response system. At a disaster-zone mobile command post, a lead-agency application coordinates triage, logistics, field assessment, communication, and hazmat agents. An edge data store receives structural, environmental, and drone data for agent use. Evacuation orders pass through HITL approval by the human incident commander, while the communication agent links the application to partner agencies and a remote command center through resilient network connectivity via low Earth orbit (LEO) satellites.}
\label{fig:emergency-response}
\end{figure*}

Consider a large-scale disaster such as an earthquake or major industrial accident, requiring coordinated response across multiple agencies. Such an emergency response situation is one of the most demanding environments for an application, as infrastructure may be damaged, decisions can carry life-or-death consequences, and multiple independent organizations must collaborate under time pressure while maintaining their own operational boundaries. The agent-native response system should be able to delegate substantial coordination and assessment work to agents that reason about unexpected situations not anticipated by predefined workflows, while keeping high-stakes decisions under human control. Fig.~\ref{fig:emergency-response} illustrates this scenario through a lead-agency mobile command post that coordinates specialized response agents, local telemetry, HITL approval, and links to partner agencies and a remote command center.

\subsubsection{System Architecture and Deployment}

The lead agency's response system is structured as a graph of specialized agents:
\begin{itemize}
  \item \textbf{Triage agent.} Processes incoming reports from responders and the public, classifies cases by severity, and prioritizes response actions.
  \item \textbf{Logistics agent.} Manages allocation of personnel, vehicles, and equipment, and coordinates their deployment to incident sites.
  \item \textbf{Field assessment agents.} Analyze drone imagery, structural sensor readings, and environmental telemetry to characterize site conditions.
  \item \textbf{Communication agent.} Coordinates with the remote command center and neighboring agencies, propagating shared situational awareness and routing inter-agency requests.
\end{itemize}
Triage findings inform logistics decisions, field assessments reshape priorities, and the communication agent propagates coordination externally. The application uses a hybrid application orchestrator where deterministic rules handle routine task assignment and escalation for human review, while an agentic component handles graph-level adaptation for novel situations such as an unexpected secondary hazard requiring the addition of a hazmat coordinator.

The system runs on edge infrastructure deployed to the disaster zone, with the following arrangements:
\begin{itemize}
  \item \textbf{Edge inference.} Servers in mobile command units host LLM inference close to where decisions are made. As no single device has the capacity to run inference for large LLMs, distributed inference splits computation across co-located edge servers, while individual responder devices run smaller specialized models.
  \item \textbf{Telemetry data delivery.} Streaming middleware routes structural-integrity, environmental, and drone-video telemetry into edge-local data stores near the agents that consume it, removing dependence on a centralized cloud that may be unreachable or have high network latency.
  \item \textbf{Resilient connectivity.} When terrestrial networks are degraded, low Earth orbit (LEO) satellites can provide wide-area connectivity or backhaul to command centers and partner agencies.
  \item \textbf{Traffic classification.} Routine status updates use best-effort network delivery, while real-time hazard alerts and physical-operation coordination can be marked for priority handling on networks that support bounded-latency service, such as TSN-capable command-post networks or URLLC-capable mobile links.
\end{itemize}

\subsubsection{Runtime Adaptation and Cross-Agency Coordination}

A novel hazard mid-incident illustrates how the agentic component in the application orchestrator can reshape the graph at runtime. Suppose that a chemical release is detected at a building already prioritized for medical evacuation, resulting in a situation beyond regular response patterns. In response, the agentic orchestrator component restructures the application graph. It spawns a hazmat-coordination agent whose assessments feed into the triage and logistics agents, so prioritization and dispatch decisions for the affected building incorporate hazmat input, including the routing of responders equipped with personal protective equipment (PPE). It also requires evacuation recommendations to pass through hazmat-aware review before final dispatch, unless a human commander overrides that requirement under emergency authority. The behavioral envelope explicitly permits this kind of change, including the type of agent that may be spawned, the allowed delegation depth, and the permission scope assigned to the new agent. The orchestrator can therefore apply the change without out-of-band reconfiguration, allowing the application graph to explore a novel response path while remaining within the declared bounds of allowable evolution.

The newly inserted hazmat coordinator soon determines that the chemical is unidentified and requires expertise the lead agency does not have. Runtime discovery locates a hazmat assessment agent at a neighboring agency, and the coordinator engages it through an agent-to-agent protocol such as A2A. The two agents communicate as peers, with each agency retaining governance control over its own agent. Fig.~\ref{fig:emergency-response} represents this pattern through links from the lead agency's communication agent to partner-agency capabilities in Agency B and Agency C.

Large-scale incident response often requires coordination among agencies with different responsibilities, such as fire services, emergency medical services, law enforcement, and hazmat teams. In an agent-native deployment, each agency may operate its own application under a separate governance domain. While agencies can reach across boundaries for capabilities they do not have internally, each agency also holds data that must not cross those boundaries, such as classified intelligence, investigation details, and personnel records. The application boundary-facing portion of each agency's agent mesh enforces these constraints, gating cross-boundary message exchanges based on agent identity, permitted interaction types, and data sensitivity.

Continuing with the incident, suppose the restructured graph converges on a recommendation to evacuate several city blocks, derived from the combined hazmat and structural assessments. Evacuation orders are designated mandatory human-approval actions in the specification. Hence the agent mesh intercepts the recommendation at the HITL checkpoint and surfaces the underlying assessment chain, policy triggers, and structured decision rationale to the human incident commander, with a bounded response window appropriate to the time pressure. The human commander approves the recommendation but expands the evacuation area by one additional city block, and execution resumes with the modified directive. Routine actions within existing permission scope, such as dispatching standard equipment or updating responder assignments, continue autonomously without HITL checkpoint overhead.

\subsubsection{Output Quality and Accountability}

While the above mechanisms control which actions the system permits, they do not establish whether the agents producing those actions are still reasoning correctly. As the incident extends over hours, the field assessment agent processing drone imagery may begin drifting. Its damage classifications may diverge from those of human assessors, while its self-reported confidence scores remain elevated. Techniques such as behavioral drift detection can help detect and flag divergence against the agent's earlier baseline, after which the application orchestrator routes pending assessments to a backup agent, while reports issued from the affected window are queued for human review.

Underlying all of these moments is an audit trail assembled from records across the runtime. Post-incident review using the audit trail can reconstruct why the application orchestrator spawned the hazmat coordinator and reshaped the graph, what the field assessment agent observed before drift was detected, which cross-agency consultations were initiated and how their outputs informed the response, and who approved each escalated decision. The same record supports accountability, regulatory review, and the improvement of response procedures across future incidents.

\subsection{Cross-Organizational Supply Chain}

\subsubsection{Application Graph and Data Foundations}

A cross-organizational supply chain scenario illustrates a different pressure point involving agents that collaborate across independent organizations with competing interests, each with its own data, policies, and strategic objectives. Consider an enterprise procurement system with a procurement agent, an analytics agent, and an inventory agent. The procurement agent evaluates supplier offers and negotiates through an agent-to-agent protocol such as A2A. The analytics agent assesses market conditions and supplier reliability, while the inventory agent tracks internal stock levels and demand forecasts. The behavioral envelope specifies which internal data each agent may access, which agents may communicate externally, and how far the procurement agent may negotiate or commit before human approval is required.

The procurement agent's effectiveness depends on access to internal data such as pricing structures, cost thresholds, inventory levels, supplier history, and strategic priorities. These data are typically distributed across organizational units including finance and warehouse operations. Data mesh principles apply naturally here because each domain team can maintain its data as a data product with quality guarantees, documentation, and discoverability, enabling agents to locate available and up-to-date data across the organization.

\subsubsection{Confidentiality and Boundary-Mediated Negotiation}

A distinctive challenge is that the procurement agent needs confidential internal information to negotiate effectively, but must not disclose that information to supplier agents. This information spans the same pricing, inventory, and strategy context that informs its internal reasoning. Data leakage can be subtle, since a message that reveals low inventory or deadline pressure may shift bargaining leverage or violate confidentiality obligations. Data access control alone is therefore insufficient, because the agent may legitimately need the information for internal reasoning. The disclosure boundary must instead be enforced by the agent mesh on outbound communication.

In this setting, each organization governs its own application graph and mesh boundary. Policy-as-code governance can define which information may be shared, potentially with contextual rules such as allowing more latitude with a trusted long-term partner than with a newly discovered supplier. When the procurement agent drafts a reply to a supplier offer, the agent mesh evaluates the outbound message against policies on information sharing. These policies can consider who will receive the message, whether it contains restricted information such as inventory levels or internal price limits, and whether the exchange is exploratory negotiation or a binding commitment. A message that says the buyer's stock of a key component is nearly exhausted, for example, can be blocked or sanitized before it reaches the supplier. Runtime discovery of a new supplier can similarly be bound into the graph only if the supplier's credentialed capability and proposed commitment scope remain inside the behavioral envelope.

\subsubsection{Commitment Oversight and Auditability}

Human oversight appears at the commitment boundary. If a supplier proposes a multi-year exclusive contract that exceeds the procurement agent's autonomy threshold, the agent mesh intercepts the recommendation and triggers a HITL review from the procurement director. The human reviewer sees the proposed terms, the supplier evidence and market analysis behind them, and the policy triggers that escalated the case. The reviewer approves, rejects, or modifies the commitment. Routine reorders within established parameters proceed autonomously. Meanwhile, the audit trail records supplier evaluations, negotiation steps, confidentiality interventions, and human inputs for later compliance review.

The agent's behavior over time also requires monitoring for procurement-specific signals. Over the course of a quarter, negotiations with newly discovered suppliers may repeatedly approach confidentiality boundaries, trigger more egress-policy interventions, or accept unusually narrow margins relative to prior supplier classes. Besides treating each event individually, the application orchestrator can also rely on the observability audit trail to analyze the agentic node's behavior over the recent period. It can interpret the accumulated pattern as signaling increased governance risk and thus tighten the agent's operating envelope, for example by lowering its autonomous commitment threshold or requiring human review for newly discovered suppliers.

\section{Conclusion}
\label{sec:conclusion}

The increasing maturity of LLM-based agents opens the way for a new application paradigm, a class of applications whose central logic is shaped by runtime agentic reasoning. Such agent-native applications have a substantially expanded space of possible behavior, including the ability to dynamically reshape the application graph itself, allowing them to respond to situations that were not enumerated at design time. At the same time, the expanded behavioral space is the source of the central governability problem, since both useful actions and unsafe or out-of-scope actions can emerge within it.

This paper proposes an architectural formulation for running such agent-native applications as graph-based systems whose runtime actions and evolution are constrained by a behavioral envelope. The envelope aims to preserve flexibility for agentic reasoning while restraining the application to a declared scope. In practice, the envelope is realized by a declarative application specification that defines both the application graph and the range of permissible actions and changes. A data-plane agent mesh and a control-plane application orchestrator apply these constraints at runtime. Assuming complete mediation, constraints that can be evaluated with sufficient reliability can be enforced independently of the behavior of individual agentic nodes. The architecture also positions HITL checkpoints and audit logging as application-level runtime responsibilities rather than obligations that each node must implement separately. The ANR prototype demonstrates a focused subset of these mechanisms in executable form.

In addition to the application model, we describe trust layer concerns spanning security, governance, and observability, as well as supporting foundations in data, cloud deployment, and networking. Lastly, two use cases illustrate how agent-native applications can operate in realistic settings.

Several limitations define the current maturity and validation scope of this work. The trace-space formulation is conceptual and does not provide complete operational semantics. The practical scope of the behavioral envelope depends on whether its policies can be evaluated reliably enough to support meaningful enforcement. ANR is a single-process feasibility prototype with simplified policies, and its audit-derived statistics do not evaluate task quality, runtime overhead, scalability, security effectiveness in production settings, or comparative performance. The use cases likewise illustrate architectural possibilities rather than validated readiness for high-stakes deployment.

Our proposal is one step toward making agent-native applications practical by addressing a substantial gap in the development of standardizable architectural mechanisms that allow such applications to be specified, run, and governed interoperably, with a particular emphasis on constraining mediated actions and runtime evolution through a behavioral envelope. Closing this gap may ultimately lead to an architectural substrate taking on a role analogous to that of Kubernetes for cloud-native applications, though the object being governed is agentic interaction and application evolution rather than container deployment. We hope the model proposed here offers groundwork for such a substrate.

\bibliography{references}

@IEEEtranBSTCTL{IEEEexample:BSTcontrol,
  CTLuse_forced_etal = "yes",
  CTLmax_names_forced_etal = "6",
  CTLnames_show_etal = "1"
}

@article{ventre2021segment_routing,
 author = {Ventre, Pier Luigi and Salsano, Stefano and Polverini, Marco and Cianfrani, Antonio and Abdelsalam, Ahmed and Filsfils, Clarence and Camarillo, Pablo and Clad, Francois},
 doi = {10.1109/COMST.2020.3036826},
 journal = {IEEE Communications Surveys \& Tutorials},
 number = {1},
 pages = {182--221},
 title = {Segment {Routing}: {A} {Comprehensive} {Survey} of {Research} {Activities}, {Standardization} {Efforts}, and {Implementation} {Results}},
 volume = {23},
 year = {2021}
}

@article{alsayyad2026agenttrace,
 author = {AlSayyad, Adam and Huang, Kelvin Yuxiang and Pal, Richik},
 publisher = {Association for the Advancement of Artificial Intelligence},
 title = {{AgentTrace}: {A} {Structured} {Logging} {Framework} for {Agent} {System} {Observability}},
 year = {2026}
}

@misc{owasp_agentic_applications,
 author = {{OWASP Gen AI Security Project, Agentic Security Initiative}},
 month = {December},
 note = {Version 2026},
 publisher = {OWASP Foundation},
 title = {{OWASP} {Top} 10 {For} {Agentic} {Applications} 2026},
 url = {https://genai.owasp.org},
 year = {2025}
}

@misc{cao2026auton,
 author = {Cao, Sheng and Chang, Zhao and Li, Chang and Li, Hannan and Fu, Liyao and Tang, Ji},
 doi = {10.48550/arXiv.2602.23720},
 note = {arXiv:2602.23720 [cs]},
 title = {The {Auton} {Agentic} {AI} {Framework}},
 year = {2026}
}

@article{wang2026internet_of_agents,
 author = {Wang, Yuntao and Guo, Shaolong and Pan, Yanghe and Su, Zhou and Chen, Fahao and Luan, Tom H. and Li, Peng and Kang, Jiawen and Niyato, Dusit},
 doi = {10.1109/TCCN.2025.3623369},
 journal = {IEEE Transactions on Cognitive Communications and Networking},
 pages = {4476--4501},
 title = {Internet of {Agents}: {Fundamentals}, {Applications}, and {Challenges}},
 volume = {12},
 year = {2026}
}

@misc{chang2025agentnetwork,
 author = {Chang, Gaowei and Lin, Eidan and Yuan, Chengxuan and Cai, Rizhao and Chen, Binbin and Xie, Xuan and Zhang, Yin},
 doi = {10.48550/arXiv.2508.00007},
 note = {arXiv:2508.00007 [cs]},
 title = {Agent {Network} {Protocol} {Technical} {White} {Paper}},
 year = {2025}
}

@inproceedings{wu2023autogen,
 author = {Wu, Qingyun and Bansal, Gagan and Zhang, Jieyu and Wu, Yiran and Li, Beibin and Zhu, Erkang and Jiang, Li and Zhang, Xiaoyun and Zhang, Shaokun and Liu, Jiale and Awadallah, Ahmed and White, Ryen W and Burger, Doug and Wang, Chi},
 booktitle = {Conference on Language Modeling (COLM)},
 title = {{AutoGen}: {Enabling} {Next}-{Gen} {LLM} {Applications} via {Multi}-{Agent} {Conversations}},
 year = {2024}
}

@inproceedings{shi2026toolhijacker,
 author = {Shi, Jiawen and Yuan, Zenghui and Tie, Guiyao and Zhou, Pan and Gong, Neil Zhenqiang and Sun, Lichao},
 booktitle = {Network and Distributed System Security ({NDSS}) {Symposium}},
 doi = {10.14722/ndss.2026.230675},
 title = {Prompt {Injection} {Attack} to {Tool} {Selection} in {LLM} {Agents}},
 year = {2026}
}

@misc{xue2024wireless_moe,
 author = {Xue, Nan and Sun, Yaping and Chen, Zhiyong and Tao, Meixia and Xu, Xiaodong and Qian, Liang and Cui, Shuguang and Zhang, Ping},
 doi = {10.48550/arXiv.2405.03131},
 note = {arXiv:2405.03131 [cs]},
 title = {{WDMoE}: {Wireless} {Distributed} {Large} {Language} {Models} with {Mixture} of {Experts}},
 year = {2024}
}

@article{zhu2022sat_terrestrial_6g,
 author = {Zhu, Xiangming and Jiang, Chunxiao},
 doi = {10.1109/JIOT.2021.3126825},
 journal = {IEEE Internet of Things Journal},
 number = {1},
 pages = {437--461},
 title = {Integrated {Satellite}-{Terrestrial} {Networks} {Toward} {6G}: {Architectures}, {Applications}, and {Challenges}},
 volume = {9},
 year = {2022}
}

@article{xu2024llm_agents_6g,
 author = {Xu, Minrui and Niyato, Dusit and Kang, Jiawen and Xiong, Zehui and Mao, Shiwen and Han, Zhu and Kim, Dong In and Letaief, Khaled B.},
 doi = {10.1109/MWC.005.2400019},
 journal = {IEEE Wireless Communications},
 number = {6},
 pages = {63--71},
 title = {When {Large} {Language} {Model} {Agents} {Meet} {6G} {Networks}: {Perception}, {Grounding}, and {Alignment}},
 volume = {31},
 year = {2024}
}

@techreport{bushaus2025modern_data_architecture,
 author = {Bushaus, Dawn and Donegan, Michelle},
 institution = {TM Forum},
 month = {August},
 title = {Unlocking {AI} potential through a modern data architecture},
 url = {https://inform.tmforum.org/research-and-analysis/reports/unlocking-ai-potential-through-a-modern-data-architecture},
 year = {2025}
}

@inproceedings{greshake2023indirect_prompt_injection,
 author = {Greshake, Kai and Abdelnabi, Sahar and Mishra, Shailesh and Endres, Christoph and Holz, Thorsten and Fritz, Mario},
 booktitle = {Proceedings of the 16th {ACM} {Workshop} on {Artificial} {Intelligence} and {Security}},
 doi = {10.1145/3605764.3623985},
 pages = {79--90},
 title = {Not {What} {You}'ve {Signed} {Up} {For}: {Compromising} {Real}-{World} {LLM}-{Integrated} {Applications} with {Indirect} {Prompt} {Injection}},
 year = {2023}
}

@misc{feng2025autonomy,
 author = {Feng, K. J. Kevin and McDonald, David W. and Zhang, Amy X.},
 doi = {10.48550/arXiv.2506.12469},
 note = {arXiv:2506.12469 [cs]},
 title = {Levels of {Autonomy} for {AI} {Agents}},
 year = {2025}
}

@misc{fleming2025layered_protocol_iot_agents,
 author = {Fleming, Charles and Pandey, Vijoy and Muscariello, Luca and Kompella, Ramana},
 doi = {10.48550/arXiv.2511.19699},
 note = {arXiv:2511.19699 [cs]},
 title = {A {Layered} {Protocol} {Architecture} for the {Internet} of {Agents}},
 year = {2025}
}

@misc{zhang2022eaas,
 author = {Zhang, Mingjin and Cao, Jiannong and Sahni, Yuvraj and Chen, Qianyi and Jiang, Shan and Wu, Tao},
 doi = {10.48550/arXiv.2209.06613},
 note = {arXiv:2209.06613 [cs]},
 title = {{EaaS}: {A} {Service}-{Oriented} {Edge} {Computing} {Framework} {Towards} {Distributed} {Intelligence}},
 year = {2022}
}

@article{zanbouri2025wireless_tsn,
 author = {Zanbouri, Kouros and Noor-A-Rahim, Md. and John, Jobish and Sreenan, Cormac J. and Vincent Poor, H. and Pesch, Dirk},
 doi = {10.1109/COMST.2024.3486618},
 journal = {IEEE Communications Surveys \& Tutorials},
 number = {4},
 pages = {2129--2155},
 title = {A {Comprehensive} {Survey} of {Wireless} {Time}-{Sensitive} {Networking} ({TSN}): {Architecture}, {Technologies}, {Applications}, and {Open} {Issues}},
 volume = {27},
 year = {2025}
}

@inproceedings{rombaut2025watson,
 author = {Rombaut, Benjamin and Masoumzadeh, Sogol and Vasilevski, Kirill and Lin, Dayi and Hassan, Ahmed E.},
 booktitle = {2025 40th {IEEE}/{ACM} {International} {Conference} on {Automated} {Software} {Engineering} ({ASE})},
 doi = {10.1109/ASE63991.2025.00067},
 pages = {739--751},
 title = {Watson: {A} {Cognitive} {Observability} {Framework} for the {Reasoning} of {LLM}-{Powered} {Agents}},
 year = {2025}
}

@techreport{chennabasappa2025llamafirewall,
 author = {Chennabasappa, Sahana and Nikolaidis, Cyrus and Song, Daniel and Molnar, David and Ding, Stephanie and Wan, Shengye and Whitman, Spencer and Deason, Lauren and Doucette, Nicholas and Montilla, Abraham and Gampa, Alekhya and de Paola, Beto and Gabi, Dominik and Crnkovich, James and Testud, Jean-Christophe and He, Kat and Chaturvedi, Rashnil and Zhou, Wu and Saxe, Joshua},
 institution = {Meta},
 month = {April},
 title = {{LlamaFirewall}: {An} open source guardrail system for building secure {AI} agents},
 url = {https://ai.meta.com/research/publications/llamafirewall-an-open-source-guardrail-system-for-building-secure-ai-agents/},
 year = {2025}
}

@inproceedings{souza2025provagent,
 author = {Souza, Renan and Gueroudji, Amal and DeWitt, Stephen and Rosendo, Daniel and Ghosal, Tirthankar and Ross, Robert and Balaprakash, Prasanna and Da Silva, Rafael Ferreira},
 booktitle = {2025 {IEEE} {International} {Conference} on {eScience} ({eScience})},
 doi = {10.1109/eScience65000.2025.00093},
 pages = {467--473},
 title = {{PROV}-{AGENT}: {Unified} {Provenance} for {Tracking} {AI} {Agent} {Interactions} in {Agentic} {Workflows}},
 year = {2025}
}

@misc{bonfim2018integrated_nfv_sdn,
 author = {Bonfim, Michel S. and Dias, Kelvin L. and Fernandes, Stenio F. L.},
 doi = {10.48550/arXiv.1801.01516},
 note = {arXiv:1801.01516 [cs]},
 title = {Integrated {NFV}/{SDN} {Architectures}: {A} {Systematic} {Literature} {Review}},
 year = {2018}
}

@inproceedings{li2019service_mesh,
 author = {Li, Wubin and Lemieux, Yves and Gao, Jing and Zhao, Zhuofeng and Han, Yanbo},
 booktitle = {2019 {IEEE} {International} {Conference} on {Service}-{Oriented} {System} {Engineering} ({SOSE})},
 doi = {10.1109/SOSE.2019.00026},
 pages = {122--127},
 title = {Service {Mesh}: {Challenges}, {State} of the {Art}, and {Future} {Research} {Opportunities}},
 year = {2019}
}

@inproceedings{rebedea2023nemo_guardrails,
 author = {Rebedea, Traian and Dinu, Razvan and Sreedhar, Makesh Narsimhan and Parisien, Christopher and Cohen, Jonathan},
 booktitle = {Proceedings of the 2023 {Conference} on {Empirical} {Methods} in {Natural} {Language} {Processing}: {System} {Demonstrations}},
 doi = {10.18653/v1/2023.emnlp-demo.40},
 pages = {431--445},
 title = {{NeMo} {Guardrails}: {A} {Toolkit} for {Controllable} and {Safe} {LLM} {Applications} with {Programmable} {Rails}},
 year = {2023}
}

@article{goedegebuure2025data_mesh_review,
 author = {Goedegebuure, Abel and Kumara, Indika and Driessen, Stefan and Van Den Heuvel, Willem-Jan and Monsieur, Geert and Tamburri, Damian Andrew and Nucci, Dario Di},
 doi = {10.1145/3687301},
 journal = {ACM Computing Surveys},
 number = {1},
 pages = {1--36},
 title = {Data {Mesh}: {A} {Systematic} {Gray} {Literature} {Review}},
 volume = {57},
 year = {2024}
}

@misc{kong2025agent_communication,
 author = {Kong, Dezhang and Lin, Shi and Xu, Zhenhua and Wang, Zhebo and Li, Minghao and Li, Yufeng and Zhang, Yilun and Peng, Hujin and Sha, Zeyang and Li, Yuyuan and Lin, Changting and Wang, Xun and Liu, Xuan and Zhang, Ningyu and Chen, Chaochao and Khan, Muhammad Khurram and Han, Meng},
 doi = {10.48550/arXiv.2506.19676},
 note = {arXiv:2506.19676 [cs]},
 title = {A {Survey} of {LLM}-{Driven} {AI} {Agent} {Communication}: {Protocols}, {Security} {Risks}, and {Defense} {Countermeasures}},
 year = {2025}
}

@misc{bhardwaj2026agent_behavioral_contracts,
 author = {Bhardwaj, Varun Pratap},
 doi = {10.5281/zenodo.18775393},
 note = {arXiv:2602.22302 [cs]},
 title = {Agent {Behavioral} {Contracts}: {Formal} {Specification} and {Runtime} {Enforcement} for {Reliable} {Autonomous} {AI} {Agents}},
 year = {2026}
}

@inproceedings{dang2025puppeteer,
 author = {Dang, Yufan and Qian, Chen and Luo, Xueheng and Fan, Jingru and Xie, Zihao and Shi, Ruijie and Chen, Weize and Yang, Cheng and Che, Xiaoyin and Tian, Ye and Xiong, Xuantang and Han, Lei and Liu, Zhiyuan and Sun, Maosong},
 booktitle = {Conference on Neural Information Processing Systems (NeurIPS)},
 title = {Multi-{Agent} {Collaboration} via {Evolving} {Orchestration}},
 year = {2025}
}

@misc{yang2025survey_agent_protocols,
 author = {Yang, Yingxuan and Chai, Huacan and Song, Yuanyi and Qi, Siyuan and Wen, Muning and Li, Ning and Liao, Junwei and Hu, Haoyi and Lin, Jianghao and Chang, Gaowei and Liu, Weiwen and Wen, Ying and Yu, Yong and Zhang, Weinan},
 doi = {10.48550/arXiv.2504.16736},
 note = {arXiv:2504.16736 [cs]},
 title = {A {Survey} of {AI} {Agent} {Protocols}},
 year = {2025}
}

@article{blohm2024data_products_mesh,
 author = {Blohm, Ivo and Wortmann, Felix and Legner, Christine and Köbler, Felix},
 doi = {10.1007/s12599-024-00876-5},
 journal = {Business \& Information Systems Engineering},
 number = {5},
 pages = {643--652},
 title = {Data products, data mesh, and data fabric: {New} paradigm(s) for data and analytics?},
 volume = {66},
 year = {2024}
}

@misc{jackson2025policy_engine,
 author = {Jackson, Freeman A.},
 howpublished = {SSRN},
 month = {November},
 title = {Designing a {Policy} {Engine} for {Agentic} {AI} {Systems}: {From} {Governance} {Requirements} to {Runtime} {Enforcement}},
 url = {https://ssrn.com/abstract=5904104},
 year = {2025}
}

@article{deng2025agent_security_threats,
 author = {Deng, Zehang and Guo, Yongjian and Han, Changzhou and Ma, Wanlun and Xiong, Junwu and Wen, Sheng and Xiang, Yang},
 doi = {10.1145/3716628},
 journal = {ACM Computing Surveys},
 number = {7},
 pages = {1--36},
 title = {{AI} {Agents} {Under} {Threat}: {A} {Survey} of {Key} {Security} {Challenges} and {Future} {Pathways}},
 volume = {57},
 year = {2025}
}

@techreport{cisco2025internet_agents,
 author = {Pandey, Vijoy},
 institution = {Outshift by Cisco},
 month = {August},
 title = {The {Internet} of {Agents}},
 url = {https://outshift.cisco.com/the-internet-of-agents},
 year = {2025}
}

@article{machado2022data_mesh,
 author = {Machado, Inês Araújo and Costa, Carlos and Santos, Maribel Yasmina},
 doi = {10.1016/j.procs.2021.12.013},
 journal = {Procedia Computer Science},
 pages = {263--271},
 title = {Data {Mesh}: {Concepts} and {Principles} of a {Paradigm} {Shift} in {Data} {Architectures}},
 volume = {196},
 year = {2022}
}

@article{finn2018tsn,
 author = {Finn, Norman},
 doi = {10.1109/MCOMSTD.2018.1700076},
 journal = {IEEE Communications Standards Magazine},
 number = {2},
 pages = {22--28},
 title = {Introduction to {Time}-{Sensitive} {Networking}},
 volume = {2},
 year = {2018}
}

@misc{wang2025mi9,
 author = {Wang, Charles L. and Singhal, Trisha and Kelkar, Ameya and Tuo, Jason},
 doi = {10.48550/arXiv.2508.03858},
 note = {arXiv:2508.03858 [cs]},
 title = {{MI9}: {An} {Integrated} {Runtime} {Governance} {Framework} for {Agentic} {AI}},
 year = {2025}
}

@article{wang2025ioa_security,
 author = {Wang, Yuntao and Pan, Yanghe and Guo, Shaolong and Su, Zhou},
 doi = {10.1109/OJCS.2025.3589638},
 journal = {IEEE Open Journal of the Computer Society},
 pages = {1611--1624},
 title = {Security of {Internet} of {Agents}: {Attacks} and {Countermeasures}},
 volume = {6},
 year = {2025}
}

@article{alonso2023multi_cloud_native,
 author = {Alonso, Juncal and Orue-Echevarria, Leire and Casola, Valentina and Torre, Ana Isabel and Huarte, Maider and Osaba, Eneko and Lobo, Jesus L.},
 doi = {10.1186/s13677-022-00367-6},
 journal = {Journal of Cloud Computing: Advances, Systems and Applications},
 number = {6},
 pages = {1--34},
 title = {Understanding the challenges and novel architectural models of multi-cloud native applications – a systematic literature review},
 volume = {12},
 year = {2023}
}

@misc{du2025agent_communication,
 author = {Du, Chenguang and Wang, Chuyi and Chao, Yihan and Xie, Xiaohui and Cui, Yong},
 doi = {10.48550/arXiv.2509.02317},
 note = {arXiv:2509.02317 [cs]},
 title = {{AI} {Agent} {Communication} from {Internet} {Architecture} {Perspective}: {Challenges} and {Opportunities}},
 year = {2025}
}

@misc{lupinacci2025inter_agent_trust,
 author = {Lupinacci, Matteo and Pironti, Francesco Aurelio and Blefari, Francesco and Romeo, Francesco and Arena, Luigi and Furfaro, Angelo},
 doi = {10.48550/ARXIV.2507.06850},
 title = {The {Dark} {Side} of {LLMs}: {Agent}-based {Attacks} for {Complete} {Computer} {Takeover}},
 year = {2025}
}

@article{guo2026agent_discovery_ioa,
 author = {Guo, Shaolong and Wang, Yuntao and Su, Zhou and Pan, Yanghe and Hu, Qinnan and Luan, Tom H.},
 doi = {10.1109/MNET.2026.3663837},
 journal = {IEEE Network},
 pages = {1--9},
 title = {Agent {Discovery} in {Internet} of {Agents}: {Challenges} and {Solutions}},
 year = {2026}
}

@article{farquhar2024semantic,
 author = {Farquhar, Sebastian and Kossen, Jannik and Kuhn, Lorenz and Gal, Yarin},
 doi = {10.1038/s41586-024-07421-0},
 journal = {Nature},
 pages = {625--630},
 title = {Detecting hallucinations in large language models using semantic entropy},
 volume = {630},
 year = {2024}
}

@inproceedings{wang2025agentspec,
 author = {Wang, Haoyu and Poskitt, Christopher M and Sun, Jun},
 booktitle = {Proceedings of the {IEEE}/{ACM} 48th {International} {Conference} on {Software} {Engineering} ({ICSE})},
 doi = {10.1145/3744916.3764546},
 title = {{AgentSpec}: {Customizable} {Runtime} {Enforcement} for {Safe} and {Reliable} {LLM} {Agents}},
 year = {2026}
}

@misc{kaptein2026governance,
 author = {Kaptein, Maurits and Khan, Vassilis-Javed and Podstavnychy, Andriy},
 doi = {10.48550/arXiv.2603.16586},
 note = {arXiv:2603.16586 [cs]},
 title = {Runtime {Governance} for {AI} {Agents}: {Policies} on {Paths}},
 year = {2026}
}

@misc{a2a_protocol_spec,
 author = {{A2A Protocol}},
 title = {Agent2Agent ({A2A}) Protocol Specification},
 url = {https://a2a-protocol.org/latest/specification/},
}

@misc{langgraph_framework,
  author = {{LangChain}},
  title = {{LangGraph}},
  url = {https://github.com/langchain-ai/langgraph},
  note = {Accessed: 2026-04-13}
}

@misc{crewai_framework,
  author = {{CrewAI, Inc.}},
  title = {{CrewAI}},
  url = {https://github.com/crewAIInc/crewAI},
  note = {Accessed: 2026-04-13}
}

@misc{openai_agents_sdk,
  author = {{OpenAI}},
  title = {{OpenAI Agents SDK}},
  url = {https://openai.github.io/openai-agents-python/},
  note = {Accessed: 2026-04-13}
}

@misc{liang2025a2h,
  author = {Liang, Zhiyuan and Cui, Enfang and Wei, Qian and She, Rui and Li, Tianzheng and Guo, Minxin and Cheng, Yujun},
  title = {{A2H}: Agent-to-Human Protocol for {AI} Agent},
  year = {2025},
  howpublished = {arXiv preprint},
  url = {https://arxiv.org/abs/2602.15831}
}

@misc{mouzouni2026context_kubernetes,
  author = {Mouzouni, Charafeddine},
  title = {Context {Kubernetes}: {Declarative} {Orchestration} of {Enterprise} {Knowledge} for {Agentic} {AI} {Systems}},
  year = {2026},
  note = {arXiv:2604.11623 [cs]},
  url = {https://arxiv.org/abs/2604.11623}
}

@misc{daunis2025declarative_agent_workflows,
  author = {Daunis, Ivan},
  title = {A {Declarative} {Language} for {Building} and {Orchestrating} {LLM-Powered} {Agent} {Workflows}},
  year = {2025},
  note = {arXiv:2512.19769 [cs]},
  url = {https://arxiv.org/abs/2512.19769}
}

@inproceedings{rao1995bdi,
  author    = {Rao, Anand S. and Georgeff, Michael P.},
  title     = {{BDI} Agents: From Theory to Practice},
  booktitle = {Proceedings of the First International Conference on Multi-Agent Systems ({ICMAS}-95)},
  year      = {1995},
  pages     = {312--319},
}

@article{kephart2003autonomic,
  author    = {Kephart, Jeffrey O. and Chess, David M.},
  title     = {The Vision of Autonomic Computing},
  journal   = {Computer},
  year      = {2003},
  volume    = {36},
  number    = {1},
  pages     = {41--50},
  doi       = {10.1109/MC.2003.1160055},
}

@misc{mcp_architecture,
  author = {{Model Context Protocol}},
  title = {Architecture Overview},
  howpublished = {Official documentation},
  url = {https://modelcontextprotocol.io/docs/learn/architecture},
  note = {Accessed: 2026-05-14}
}

@misc{lfedge_edgelake,
  author = {{LF Edge}},
  title = {{EdgeLake}},
  howpublished = {Project page},
  url = {https://lfedge.org/projects/edgelake/},
  note = {Accessed: 2026-05-02}
}

@misc{nvidia_openshell_overview,
  author = {{NVIDIA}},
  title = {Overview of {NVIDIA OpenShell}},
  howpublished = {Official documentation},
  url = {https://docs.nvidia.com/openshell/latest/about/overview},
  note = {Accessed: 2026-06-04}
}

@inproceedings{varghese2017edge_service,
 author = {Varghese, Blesson and Wang, Nan and Li, Jianyu and Nikolopoulos, Dimitrios S.},
 booktitle = {International Conference on Parallel Computing},
 pages     = {784--793},
 title = {Edge-as-a-Service: Towards Distributed Cloud Architectures},
 year = {2017}
}

@inproceedings{morabito2016edge_iot,
 author = {Morabito, Roberto and Beijar, Nicklas},
 booktitle = {2016 {IEEE} {International} {Conference} on {Sensing}, {Communication} and {Networking} ({SECON} {Workshops})},
 doi = {10.1109/SECONW.2016.7746807},
 pages = {1--6},
 title = {Enabling {Data} {Processing} at the {Network} {Edge} through {Lightweight} {Virtualization} {Technologies}},
 year = {2016}
}

@article{li2021edge_leo,
 author = {Li, Chengcheng and Zhang, Yasheng and Xie, Renchao and Hao, Xuekun and Huang, Tao},
 doi = {10.1109/ACCESS.2021.3064397},
 journal = {{IEEE} {Access}},
 pages = {39126--39137},
 title = {Integrating {Edge} {Computing} into {Low} {Earth} {Orbit} {Satellite} {Networks}: {Architecture} and {Prototype}},
 volume = {9},
 year = {2021}
}

@article{he20216g_cav,
 author = {He, Jianhua and Yang, Kun and Chen, Hsiao-Hwa},
 doi = {10.1109/MNET.011.2000541},
 journal = {{IEEE} {Network}},
 number = {4},
 pages = {255--261},
 title = {{6G} {Cellular} {Networks} and {Connected} {Autonomous} {Vehicles}},
 volume = {35},
 year = {2021}
}

@misc{openai2026codex_long_horizon,
  author = {Choi, Derrick},
  title = {Run Long Horizon Tasks with {Codex}},
  howpublished = {{OpenAI Developers} blog post},
  url = {https://developers.openai.com/blog/run-long-horizon-tasks-with-codex},
  month = {February},
  year = {2026},
  note = {Accessed: 2026-06-06}
}

@misc{anthropic2026long_running_claude_scientific,
  author = {Mishra-Sharma, Siddharth},
  title = {Long-Running {Claude} for Scientific Computing},
  howpublished = {{Anthropic} research post},
  url = {https://www.anthropic.com/research/long-running-Claude},
  month = {March},
  year = {2026},
  note = {Accessed: 2026-06-06}
}

@misc{anthropic2026parallel_claudes_c_compiler,
  author = {Carlini, Nicholas},
  title = {Building a {C} Compiler with a Team of Parallel {Claudes}},
  howpublished = {{Anthropic} engineering post},
  url = {https://www.anthropic.com/engineering/building-c-compiler},
  month = {February},
  year = {2026},
  note = {Accessed: 2026-06-06}
}

@misc{cursor2026scaling_agents,
  author = {Lin, Wilson},
  title = {Scaling Long-Running Autonomous Coding},
  howpublished = {{Cursor} research blog post},
  url = {https://cursor.com/blog/scaling-agents},
  month = {January},
  year = {2026},
  note = {Accessed: 2026-06-06}
}

@article{buehler2026agentbound,
  author = {B{\"u}hler, Christoph and Biagiola, Matteo and Di Grazia, Luca and Salvaneschi, Guido},
  title = {{AgentBound}: Securing Execution Boundaries of {AI} Agents},
  journal = {Proceedings of the ACM on Software Engineering},
  year = {2026},
  doi = {10.1145/3808103},
}

@misc{wang2026sandlock,
  author = {Wang, Cong and Zheng, Yusheng},
  title = {{Sandlock}: Confining {AI} Agent Code with Unprivileged {Linux} Primitives},
  howpublished = {arXiv preprint arXiv:2605.26298},
  year = {2026},
  doi = {10.48550/arXiv.2605.26298}
}

@misc{twingate2025identity_firewall,
  author = {Liu, Anna},
  title = {Introducing {Identity Firewall}: Security, Performance, Simplicity. Pick Three.},
  howpublished = {{Twingate} blog post},
  url = {https://www.twingate.com/blog/identity-firewall},
  month = {November},
  year = {2025},  
  note = {Accessed: 2026-06-07}
}

@misc{equinix2026data_problem_agentic_ai,
  author = {Tairych, David},
  title = {The Data Problem Behind Agentic {AI}, and What You Can Do About It},
  howpublished = {{Equinix} blog post},
  url = {https://blog.equinix.com/blog/2026/01/29/the-data-problem-behind-agentic-ai-and-what-you-can-do-about-it/},
  month = {January},
  year = {2026},
  note = {Accessed: 2026-06-08}
}

\noindent\begin{minipage}[t]{\columnwidth}
\vspace*{-3\baselineskip}

\begin{IEEEbiography}[
  {\includegraphics[
    width=1in,
    height=1.25in,
    clip,
    keepaspectratio
  ]{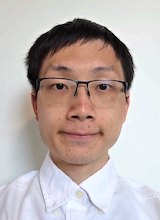}}
]{David Luo}
is a MASc student in the Department of Electrical and Computer
Engineering at the University of Toronto, where he previously
received his BASc degree. He is also an AI architect at
DodonaData.ai, working on building agentic systems for enterprise
use. His research interests include LLM agents, evaluation of
LLM-based systems, and cloud and networking infrastructure for AI
workloads.
\end{IEEEbiography}

\begin{IEEEbiography}[
  {\includegraphics[
    width=1in,
    height=1.25in,
    clip,
    keepaspectratio
  ]{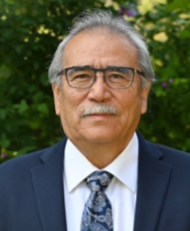}}
]{Alberto Leon-Garcia}
is Professor in Electrical and Computer Engineering at the
University of Toronto. He is a Fellow of the Royal Society of
Canada and a Life Fellow of the Institute of Electrical and
Electronics Engineers. Leon-Garcia was Scientific Director of the
NSERC Strategic Network for Smart Applications on Virtual
Infrastructures (SAVI) and Principal Investigator of the project
on Connected Vehicles and Smart Transportation (CVST). SAVI
designed and deployed a national testbed in Canada that converges
cloud computing and software-defined networking. CVST designed
and deployed an application platform for smart transportation.
Leon-Garcia's research is focussed on the design of intelligent
automated management systems for large-scale infrastructures such
as telecom networks and safe city platforms. Leon-Garcia was
Founder of AcceLight Networks and StreamWorx.AI, and in 2025 he
launched DodonaData.ai. He wrote \emph{Probability and Random
Processes for Electrical Engineering} and \emph{Communication
Networks: Fundamental Concepts and Key Architectures}, which have
been used by students around the world.
\end{IEEEbiography}

\EOD
\end{minipage}

\end{document}